\documentclass[11pt]{article}
\usepackage[english]{babel}
\usepackage[margin=1in]{geometry}
\usepackage[english]{babel}
\usepackage[utf8]{inputenc}
\usepackage{johd}
\usepackage{wasysym}
\usepackage{amsthm, amssymb}
\usepackage{circledsteps}

\newtheorem{result}{Result}
\usepackage{graphicx}
\usepackage{float}
\usepackage{natbib}
\usepackage{booktabs}
\usepackage{appendix}
\usepackage[flushleft]{threeparttablex}
\usepackage{ragged2e}
\usepackage{caption}
\usepackage{subfig}
\usepackage{hyperref}
\usepackage{setspace}
\usepackage{multirow}
\usepackage{pifont}
\newcommand{\cmark}{\ding{51}}%
\newcommand{\xmark}{\ding{55}}%

\usepackage{color}

\newcommand\Flat{\textit{\scshape{Flat}}}
\newcommand\Progressive{\textit{\scshape{Prog}}}
\newcommand\Regressive{\textit{\scshape{Reg}}}
\newcommand\AI{\textit{\scshape{PostAI}}}
\newcommand\NoAI{\textit{\scshape{PreAI}}}

\title{Overcoming Medical Overuse with AI Assistance: \\ An Experimental Investigation\footnotemark[1]}

\author{Ziyi Wang$^{a}$, Lijia Wei$^{a}$$^{*}$, Lian Xue$^{a}$ \\
        \small $^{a}$Economics and Management School, Wuhan University, Wuhan, China \\
        \small $^{*}$All authors contributed equally; Corresponding author email: \tt{ljwei@whu.edu.cn} \\
}

\date{}

\begin{document}
\maketitle
\begin{abstract} 
\noindent 

This study evaluates the effectiveness of Artificial Intelligence (AI) in mitigating medical overtreatment, a significant issue characterized by unnecessary interventions that inflate healthcare costs and pose risks to patients. We conducted a lab-in-the-field experiment at a medical school, utilizing a novel \emph{medical prescription task}, manipulating monetary incentives and the availability of AI assistance among medical students using a three-by-two factorial design. We tested three incentive schemes: Flat (constant pay regardless of treatment quantity), Progressive (pay increases with the number of treatments), and Regressive (penalties for overtreatment) to assess their influence on the adoption and effectiveness of AI assistance. Our findings demonstrate that AI significantly reduced overtreatment rates—by up to 62\% in the Regressive incentive conditions where (prospective) physician and patient interests were most aligned. Diagnostic accuracy improved by 17\% to 37\%, depending on the incentive scheme. Adoption of AI advice was high, with approximately half of the participants modifying their decisions based on AI input across all settings. For policy implications, we quantified the monetary (57\%) and non-monetary (43\%) incentives of overtreatment and highlighted AI's potential to mitigate non-monetary incentives and enhance social welfare. Our results provide valuable insights for healthcare administrators considering AI integration into healthcare systems.
\end{abstract} 

\noindent\\

\begin{quote}
\renewcommand{\thefootnote}{*}
\footnotetext[1]{We thank Mengyuan Wu, Shapeng Jiang, Yanyu Li, Xiangyu Gao, Qiuhao Chen, Wenjun Li for their assistance with the experimental sessions. Lijia Wei acknowledges support from National Science Foundation China (72173093), and the Center for Behavioral and Experimental Research (CBER) at Wuhan University. Lian Xue acknowledges support from the Research Funds for Youth Academic Team in Humanities and Social Sciences of Wuhan University (413000425). The authors declare no conflict of interest.}

\emph{``The psychologist Gerd Gigerenzer has a simple heuristic. Never ask the doctor what you should do. Ask him what he would do if he were in your place. You would be surprised at the difference''}

\hfill --- \emph{Nassim Nicholas Taleb, Antifragile: Things That Gain from Disorder}
\end{quote}

\section{Introduction}

Overtreatment in healthcare, defined as unnecessary medical interventions that can harm patients, represents a significant economic and health concern. It leads to billions of dollars in wasted spending and exposes patients to unnecessary physical and psychological risks \citep{muskens2022overuse}. This phenomenon is well-documented in both the clinical and health economics literature. Examples of overtreatment include the excessive use of antibiotics for viral infections --medical overuse or overprescription-- and the overuse of imaging tests, such as CT scans and MRIs, for conditions that could be managed with less invasive methods \citep{morgan2015setting}. These practices not only strain healthcare budgets but also expose patients to unnecessary physical and psychological risks. A recent Lancet study highlights the pervasiveness of overtreatment, attributing 29\% of healthcare spending in the US to it, with costs reaching as high as 89\% in certain populations globally, and rising in low- and middle-income countries \citep{wennberg2002geography,korenstein2012overuse,brownlee2017evidence,albarqouni2023overuse}.

A major hurdle in studying overtreatment is the difficulty of measuring overuse accurately \citep{brownlee2017evidence}. Existing methodologies employed both direct and indirect measures. Direct approaches involve using patient registries and medical records guided by evidence-based guidelines or expert consensus panels, such as the RAND Appropriateness Method \citep{fitch2000rand}. These require a precise definition of the ``appropriate care'', which often lacks clarity in many clinical settings. Furthermore, guidelines typically do not provide the necessary details for individual patient care, challenging the case-specific flexibility in measuring overtreatment \citep{chassin1998urgent,korenstein2012overuse}. Indirect approaches involve the identification of unexpected variations in healthcare service utilization both within and across countries in the healthcare system. For example, excessive use of insulin --compared to the means-- for patients with diabetes could be considered as medical overuse \citep{brownlee2017evidence}. However, these methods struggle with standardization and often yield inconclusive results.

This paper presents a novel experimental approach designed to measure the medical overuse tendencies of prospective physicians (medical students) from a behavioral perspective. We construct a virtual doctor-patient consultation scenario, using cases from standard practice exam questionnaires that simulate the patient's illness presentation. Given the description of the illness scenario, participants were instructed to select the most appropriate medicine from a set of five options. They were informed that there was only one correct answer but could choose up to two responses (medicines) for each question (illness). To incentivize accurate diagnosis and mimic real-world interactions, each choice is associated with payoffs for both the participant (physician) and a stylised patient.
Specifically, participants' choices are linked to a donation to a patient-regarding charity, where overuse of medicine is associated with a lower amount of donation compared to one exact correct treatment. We refer to this task as the \emph{medical prescription task}. This setup has several advantages and complements existing measurements of medical overuse: it can be easily generalized to all medical services with category-specific questions; it is anonymous and harmless, with respect to physicians' reputations and patients' well-being.

Utilizing the medical prescription task framework in randomized controlled trials, we aim to disentangle the underlying motives for physicians' propensity to overuse medicines (\emph{causes}) and explore effective methods to curb medical overuse (\emph{interventions}), which represent the two primary objectives of this research. The causes of medical overuse can be summarized into two categories: monetary and non-monetary. Monetary incentives, such as fee-for-service payment models, often encourage physicians to the volume of care delivered rather than its quality \citep{pauly2012handbook,morgan20192019,lu2014insurance}. Non-monetary incentives may involve defensive medicine, where doctors prescribe unnecessary treatments to avoid litigation \citep{summerton1995positive}, a lack of adequate knowledge, leading to decisions that do not align with the latest clinical guidelines \citep{price1986doctors,bishop2010physicians,jatoi2019clinical}, and out of demand pressure from patients when there is price reduction in the medicines \citep{lopez2018contribution}. 
Systematic reviews indicate that medical overuse and its underlying incentives are deeply embedded within healthcare systems, necessitating further exploration of identification and intervention methods \citep{korenstein2012overuse,tung2018factors,segal2022factors}.

In our experiment, we exogenously vary three types of incentive schemes for prospective physicians: a \emph{flat} scheme where participant payoff remains unaffected by the choices made in the medical prescription task; a \emph{progressive} scheme,  where payoff increases with the number of choices selected; and a \emph{regressive} scheme where overtreatment resulted in reduced payoff, contingent upon whether the correct medicine was selected. Patients' payoffs remained constant across all three treatments, with their welfare maximized when exactly one correct medicine was selected by the (prospective) physician (\cmark), followed by `one correct plus one incorrect medicine' (\cmark\xmark). `One incorrect' (\xmark) and `two incorrect' (\xmark\xmark) choices progressively decreased welfare. Notably, in the Regressive treatment, physicians' payoffs followed the same pattern as patients'. In this case, we say that physicians and patients have \emph{aligned interests}, mirroring the quote from `Antifragile' at the beginning of this paper \citep{taleb2014antifragile}.

Additionally, we explore the impact of AI decision-support on physician choices within each treatment. For each medical prescription task, participants made an initial decision on their own. Subsequently, they were presented with a descriptive analysis labelled ``AI diagnosis'' along with a specific treatment suggestion termed ``AI choice'' before making a second choice. This AI advice --diagnosis and choice-- was generated in advance using the large language model, ChatGPT 4.0, which achieved an accuracy rate of 73.93\% across more than 800 evaluations \citep{openai}. In real life, AI has the potential to significantly mitigate overtreatment by offering evidence-based recommendations that enhance clinical decision-making \citep{jiang2017artificial,topol2019high}. AI systems can analyze vast amounts of data to identify the most appropriate treatments, potentially helping to avoid unnecessary medical procedures \citep{pine2012harnessing,obermeyer2016predicting}. Existing literature on AI-assisted decision-making has demonstrated that these technologies can enhance the accuracy of diagnoses and tailor treatment plans to individual patients \citep{niel2019artificial, davenport2019potential,sung2024artificial}.
However, implementing AI in healthcare still faces challenges such as ethical and equity concerns, patient's trust towards AI diagnosis and physicians' acceptability of algorithms \citep{dietvorst2015algorithm,char2018implementing, emanuel2019artificial,glauser2020ai,al2023review}. Our paper contributes by providing the first experimental evidence of the effectiveness of AI on overtreatment and corresponding welfare consequences.

Our results were centred around two key performance measures: the quantity of prescribed medicines and the quality of task accuracy. 
For quantity measures, exploiting the exogenous variations in incentive schemes and the provision of AI assistance in medical prescription tasks, we could identify both the monetary (incentive schemes) and non-monetary (AI-enhanced decision accuracy and others) influences on overtreatment. We observed that overtreatment rates among participants ranged from 24-58\% across all treatments. Further analysis revealed that more than half (57\%) of the overtreatment was due to monetary incentives, which can be reduced when overtreatment is penalized with a regressive incentive scheme. AI significantly reduce overtreatment in all settings, with the most pronounced effect observed in the regressive treatment, achieving a reduction rate of 62\%. We refer to this phenomenon as the \emph{Synergistic AI Effect}, highlighting how AI's effectiveness in reducing medical overuse is maximized when physicians and patients have aligned interests.

Regarding quality, we found that a progressive incentive encouraging more prescriptions could increase accuracy because participants included more choices in their answers compared to a fixed payment-- i.e., overuse as a conservative treatment. The introduction of AI significantly boosted the accuracy rate across all treatments, with the largest effect in the Flat treatment with constant payoffs --an enhancement rate of 37\%.%
\footnote{\citet{almog2024ai} finds that umpires lowered their overall mistake rate after the introduction of Hawk-Eye review in top tennis tournaments. The AI assistance in our paper is functionally similar to Hawk-Eye.}
Intriguingly, the positive impact of AI can be further differentiated into two distinct channels: an increase in the provision of `one exact medicine \cmark' (termed the \emph{Optimal Treatment Adjustment}) and the refinement of `one correct plus one incorrect medicine \cmark\xmark' (referred to as the \emph{Precision-Compromise Strategy}). These effects vary significantly across different incentive schemes, with the regressive treatment most favoring the Optimal Treatment Adjustment over the Precision-Comparomise Strategy. In other words, the effect of AI on reducing \emph{redundant inaccurate} prescriptions is most pronounced in the regressive treatment.

This study enhances the existing body of research by empirically examining the effectiveness of AI interventions in curbing overtreatment in healthcare decisions. By analyzing overtreatment and accuracy rates before and after the deployment of AI recommendations, our findings offer robust evidence supporting AI's role in minimizing unnecessary medical practices and improving diagnosis accuracy \citep{kelly2019key,chen2023framework}. Furthermore, this research advances our comprehension of AI's potential integration into clinical workflows, thereby improving decision-making processes \citep{wang2019ai,recht2020integrating}. Specifically, it aligns with the literature that discusses the modification of physician incentives, suggesting that AI can mitigate the incentive-driven overtreatment in flat or pay-for-service incentives (progressive treatment) and an advanced effect in the regressive incentive scheme, thus addressing issues related to incentive-driven overutilization \citep{clemens2014physicians,scott2002incentives,ettner2012role,xi2019does,vlaev2019changing}. Additionally, it contributes to digital healthcare literature by demonstrating how technology-driven solutions can effectively supplement traditional healthcare practices, leading to more efficient and patient-centered care \citep{sheikh2021health, mesko2017digital,iqbal2022regulatory,ibeneme2022strengthening}.

Policy considerations should prioritize exploring the interaction effect of AI and incentive schemes on physician behavior. This research highlights the potential for AI to mitigate overtreatment, and its effectiveness could be further amplified by aligning physician incentives with AI adoption. Policymakers can explore alternative payment models that reward physicians who utilize AI tools and make treatment decisions aligned with AI recommendations and patients' welfare, particularly when such recommendations discourage overtreatment. Furthermore, given the challenges of implementing new incentive schemes in the healthcare industry \citep{agrawal2022power}, our results suggest an alternative intervention that is \emph{as effective as} introducing a pay-by-performance (\Regressive{}) incentive scheme at ensuring healthcare quality -- i.e., to incorporate AI within existing flat (pay-by-visit) incentive structure. Finally, our experiment provides an example of clinician-AI collaboration, which advocates policies that foster trust and ensure that AI complements, rather than replaces, physician expertise.

The remainder of this paper is organized as follows: Section~\ref{sec:design} presents the experimental design and procedures; Section~\ref{sec:result} reports results, and Section~\ref{sec:conclusion} concludes with a discussion.

\section{Experimental Design and Procedure}
\label{sec:design}

We conducted a lab-in-the-field experiment at a medical school affiliated with Central South Hospital, utilizing a two-by-three factorial design. Participants in this study were prospective physicians recruited from this institution, where they are being trained as medical doctors.

\paragraph{Incentive Schemes (Between-Subject)}
The experiment varied the incentive structures for participants during the main task. We tested three types of incentives:
 \Flat{}, \Progressive{}ressive, and \Regressive{}ressive in a between-subject design. 

\paragraph{AI Assistance (Within-Subject)}
We also varied the decision-making process by introducing an AI assistance feature. Participants made initial choices independently and subsequently could remake their decisions after viewing AI-generated recommendations. The AI advice, powered by ChatGPT 4.0, included both a specific recommendation and an extended explanation of the optimal medication choice based on patient symptoms. 

\paragraph{Control Variables}
To account for individual fixed effects and other potential confounders affecting the propensity for overtreatment, we implemented pre- and post-experiment assessments. These assessments evaluated various attributes, including physicians' professional abilities, cognitive reflection \citep{frederick2005cognitive}, cognitive ability \citep{carpenter1990one}, risk preference \citep{blais2006domain},  altruism preference and trust \cite{falk2018global}, algorithm literacy, trust in algorithms, awareness of algorithms, and perceptions of algorithm fairness.%
\footnote{We also conducted a standard array of questionnaires, including gender, major of study, age, grade, and education. Instructions for the experiment and for the control question are available in the Appendix~\ref{sec:instruction}. }

\subsection{Medical Prescription Task}
Participants were asked to answer \emph{20} multiple-choice questions that simulated a real-life doctor-patient consultation, with basic patient information and disease descriptions from the ``China's National Qualifying Examination for Medical Practitioners''
\footnote{Medical students in China need to obtain the ``National Practicing Physician Qualification Certificate"  through the examination before they are qualified to practice medicine and can officially join the profession.}
question bank, which consists of 817 questions. Each question had only one correct answer, but participants could select up to two answers to simulate potential overtreatment scenarios. Feedback on their accuracy was withheld until the end of the experiment. An example of such a question is provided below:

\begin{quote} 

    \textit{Female, 54 years old, with a history of hyperthyroidism. Recently, due to overwork and emotional stress, she has experienced insomnia, and heart and chest discomfort. Physical examination: Heart rate 160 beats per minute, electrocardiogram shows clear signs of myocardial ischemia, and sinus rhythm irregularity. The best choice would be:}
    
    \textit{A. Amiodarone \\
    B. Quinidine \\
    C. Procainamide \\
    D. Propranolol \\
    E. Lidocaine }
\end{quote}

To examine participants' innate ability at the medical prescription task, they were instructed to complete a separate set of 10 medical prescription tasks as a practice at the beginning of the experiment, with each correct choice yielding a 1 yuan bonus. 

\subsection{Treatment Design}

We vary how participants were incentivized and whether AI assistance was provided during the medical prescription task. The treatments regarding incentive schemes were structured as follows:

\begin{itemize}
    \item \textbf{\Flat{}}: Participants received a constant monetary payoff of 3 yuan for each medial prescription task completed, regardless of their choice.
    \item \textbf{\Progressive}ressive: Participants received a higher payoff for selecting two options (4 yuan) versus only one (2 yuan) option, irrespective of correctness.
    \item \textbf{\Regressive}ressive: Payoffs varied depending on the accuracy and number of choices: 6 yuan for one correct choice (\cmark), 4.5 yuan for two choices including the correct one (\cmark\xmark), 1.5 yuan for one incorrect choice (\xmark), and 0 yuan for two incorrect choices (\xmark\xmark).    
\end{itemize}

\noindent{\textbf{Patients' Payoffs. }} Upon successful completion of each medical prescription task, we donated the patients' payoff generated by the participants to a patient-regarding charity, specifically the Tencent Public Welfare ``Love Angel'' project, which supports children suffering from leukemia and cancer. The payoff structure for the patients was designed as follows: a base payoff of $P$ yuan was established, with an increase of $B$ yuan for correctly prescribed medicine and a decrease of $C$ yuan for incorrectly prescribed medicine. For the purposes of our experiment, the values were set at $P = 4$, $B = 4$, and $C = 2$. Notably, in the Regressive treatment, physicians' and patients' payoffs were exactly collinear. This means that physicians made decisions for patients as if they were making decisions for themselves from a payoff perspective.  The detailed experimental design, including incentive schemes and patients' payoffs, is summarized in Table~\ref{tab:design}.

\begin{table}[h!]
\caption{Experimental Design: Incentive Schemes and Patients' Payoff}
\centering
\begin{threeparttable}
\setlength{\tabcolsep}{1mm}{
\begin{tabular}{lcccccc} 
\toprule
 \multirow{2}{*}{\textbf{Choice}}  & \multicolumn{3}{c}{\textbf{Physicians' Payoff}} && \multicolumn{2}{c}{\textbf{Patients' Payoff}} \\ 
\cmidrule{2-4} \cmidrule{6-7}
& \Flat{} & \Progressive ressive & \Regressive ressive && Amount & Calculation \\
\midrule
Correct (\cmark) & 3 & 2 & 6 && 8 & \(P + B = 4 + 4\) \\
Correct/Incorrect (\cmark\xmark) & 3 & 4 & 4.5 && 6 & \(P + B - C = 4 + 4 - 2\) \\
Incorrect (\xmark) & 3 & 2 & 1.5 && 2 & \(P - C = 4 - 2\) \\
Double Incorrect (\xmark\xmark) & 3 & 4 & 0 && 0 & \(P - 2C = 4 - 2 \times 2\) \\
\bottomrule    
\end{tabular}}
\begin{tablenotes}
\small
\item \textit{Notes:} This table reports the incentive schemes for physicians and the corresponding payoffs for patients based on the accuracy and number of choices made. The symbol \cmark{} denotes the correct choice, while \xmark{} denotes an incorrect one. The parameters $P$, $B$, and $C$ represent the base payoff, bonus for a correct prescription, and penalty for an incorrect prescription, respectively, each measured in yuan.
\end{tablenotes}
\end{threeparttable}
\label{tab:design}
\end{table}

\noindent\textbf{AI assistance. }Participants made the choice twice in the medical prescription task: an initial choice without AI and a second choice with AI assistance. The AI advice, powered by ChatGPT 4.0 in early 2024, entails two parts: an exact recommendation of the optimal medication choice given the patient's symptoms and an extended explanation of the recommended choice. An example of the AI assistance is as follows:
\begin{quote} \textit{
    The AI recommends option: D. \\
    Analysis: Based on the patient's symptoms and physical examination results, the optimal medication choice is usually D. Propranolol. This beta-blocker is commonly used to control rapid heart rates and arrhythmias, particularly in hyperthyroidism cases. It helps slow down the heart rate, reduce the cardiac workload, and alleviate myocardial ischemia symptoms. However, a specific treatment plan should be consulted with a physician.}
\end{quote}
To best incentivize participants' effort at both the initial and revised decisions, participants' payoffs were determined by either their initial or second choice, randomized at the session level by the computer.

\subsection{Procedures}
The experiment was conducted at the affiliated medical school of Central South Hospital. A total of 120 medical students were recruited for the study, with 40 students participating in each of the three treatment groups. These students were part of the subjects pool for the Center for Behavioral and Experimental Research (CBER) from Wuhan University. Upon arrival, participants were allocated to one of two seminar rooms on campus, with each session comprising 5 participants. Each session lasted between 50 to 60 minutes. All tasks were computerized using the oTree platform \citep{chen2016otree}. The average payoffs were 104.7 yuan: 95.9 yuan for the \Flat{}, 93.8 yuan for the \Progressive{}, and 124.3 yuan for the \Regressive{} treatment respectively.

\section{Results}
\label{sec:result}

\subsection{Overview}

This section outlines the two key outcome variables that assess the performance of doctors in terms of \emph{quantity} and \emph{quality}.

\paragraph{Quantity}
We define \emph{overuse} as the selection of two answers in each medical prescription task. Given that there is a single correct answer and participants were fully aware of this fact, any response with more than one answer is categorized as overtreatment.

\paragraph{Quality}
Our main measurement of quality is the accuracy of the diagnosis, which hinges on whether participants chose the correct answer. Without loss of generality, let us denote the correct choice as \( A \). The \emph{accuracy rate} is then defined as the proportion of choices that include at least one \( A \). This can be mathematically expressed as:
\begin{equation}
    Y = A \cup (A \times \complement_{A}) 
    \label{eq:accuracy}
\end{equation}
where \( \complement_{A} \equiv U \setminus A = B \cup C \cup D \cup E \), and \( U \) represents the universal set of all possible choices.

\bigskip
In addition to the performance measures, we examine the extent to which participants adopt or trust AI advice subsequent to their initial choices. Participants initially made a choice without AI guidance, after which they were presented with the AI recommendation. Subsequently, they were prompted to make a second choice. Compensation for the session was determined randomly, based on either the first or second choice, as selected by the computer.

\paragraph{AI adoption} is characterized by whether participants altered their choices following the AI advice. We employ two measures of AI adoption: a flexible measure and a more restrictive measure. Under the flexible measure, maintaining the initial answer despite AI's recommendation to change is classified as \emph{not following} AI advice, whereas all others are considered \emph{following}. For instance, if the first choice was \( B \), AI suggested \( A \), and the second choice remained \( B \), this is deemed as not following AI advice. Conversely, any other alteration is categorized as following.
The more restrictive measure defines only \emph{active} changes that align with AI advice as \emph{AI adoption}. All other instances are considered \emph{non-adoption}. For example, a sequence of $B-A-B$ is defined as not following AI advice as with the main measurement. However, a sequence of $A-A-A$ is defined as not following since there are no active changes in choices.
Table~\ref{tab:ai_adopt_def} presents an exhaustive list of scenarios and their corresponding classifications regarding AI adoption. Table~\ref{tab:overview} reports the summary statistics of the three key outcome variables before and after AI intervention across treatments.

\begin{table}[h!]
\centering
\caption{Two measurements of AI adoption (Within-Subject)}
\setlength{\tabcolsep}{1mm}{
\begin{tabular}{cccccc}
\toprule
\multicolumn{3}{c}{\textbf{Decision Process}} & \multicolumn{3}{c}{\textbf{AI Adoption Measures}} \\ 
\cmidrule(lr){1-3} \cmidrule(lr){4-6}
First Choice & AI Advice & Second Choice & Main & \textit{Adjusted} & \textit{Prop (\%)} \\ 
\midrule
$A \cup (A \times \complement_{A})$ & A & $A \times \complement_{A}$ & 1 & 0 & 48.5 \\
$A \times \complement_{A}$ & A & $A$ & 1 & 1 & 13.8 \\
$B \cup (B \times \complement_{A \cup B})$ & A & $A \cup (A \times \complement_{A})$ & 1 & 1 & 34.3 \\
$B \cup (B \times \complement_{B})$ & A & $B \cup (B \times \complement_{A\cup B}) \cup C \cup (C \times \complement_{A\cup B})$ & 0 & 0 & 3.5 \\
\bottomrule
\end{tabular} }
\begin{tablenotes}
    \small
    \item \textit{Notes:} This table provides an exhaustive combination of decision-making scenarios before and after AI intervention and measures AI adoption based on two metrics. In the `Main' and `Adjusted' columns, a `1' indicates adoption of AI advice, and `0' indicates non-adoption. The key difference between the two measurements is whether an unchanged choice that is consistent with AI advice is considered as following AI. For example, AC-A-AC is considered as following in main but not following AI advice in adjusted measurements (first-row). Whereas, AC-A-A is considered as following AI advice in both measurements (second row). In the last column, we listed the overall proportion of choices under each scenario to facilitate readers' own judgement. 
\end{tablenotes}
\label{tab:ai_adopt_def}
\end{table}

\begin{table}[h!]
    \centering
    \caption{Summary Statistics}
    \begin{threeparttable}
    \begin{tabular}{lcccccc}
    \toprule
    \multirow{2}{*}{\textbf{Sample}} & \multirow{2}{*}{\textbf{Overuse (\%)}} & \multirow{2}{*}{\textbf{Accuracy Rate (\%)}} & \multicolumn{2}{c}{\textbf{AI Adoption (\%)}} \\
    \cmidrule(lr){4-5}
    & & &Main & \textit{Adjusted} \\
    \midrule
    \multicolumn{5}{l}{\textit{Panel A. Flat Treatment}} \\
    \NoAI{} & 24.3 & 56.3 & - & - \\
    \AI{} & 16.0 & 77.3 & - & - \\
    Total & 20.1 & 66.8 & 95.3 & 48.9 \\
    $p$-value & $0.004$ & $<0.001$ & - & - \\
    \midrule
    \multicolumn{5}{l}{\textit{Panel B. Progressive Treatment}} \\
    \NoAI{} & 58.1 & 67.3 & - & - \\
    \AI{} & 42.9 & 82.5 & - & - \\
    Total & 50.5 & 74.9 & 97.6 & 47.3 \\
    $p$-value & $0.001$ & $<0.001$ & - & - \\
    \midrule
    \multicolumn{5}{l}{\textit{Panel C. Regressive Treatment}} \\
    \NoAI{} & 26.0 & 66.1 & - & - \\
    \AI{} & 9.8 & 77.5 & - & - \\
    Total & 17.9 & 71.8 & 96.8 & 48.1 \\
    $p$-value & $<0.001$ & $<0.001$ & - & - \\
    \bottomrule
    \end{tabular}
    \begin{tablenotes}
        \small
        \item \textit{Notes:} This table reports summary statistics for overuse and accuracy rates across treatments, before (\NoAI) and after AI intervention (\AI). Additionally, it reports AI adoption percentages post-intervention. The columns labelled ``Main'' and ``Adjusted'' refer to different AI adoption metrics defined in Table~\ref{tab:ai_adopt_def}. $p$-values, derived from Wilcoxon signed-rank tests, assess statistical significance in changes before and after AI intervention.
    \end{tablenotes}
    \end{threeparttable}
    \label{tab:overview}
\end{table}

\subsection{The effect of AI on medical overuse}

\begin{result}[AI on Overuse]
\label{res:overuse} 
    The use of AI can significantly reduce the propensity of medical overuse among medical students, with an overall reduction effect of 37\% across all treatments. This effect is the most pronounced in the \Regressive{} treatment.
\end{result}

\begin{proof}[Support.]
Support for Result~\ref{res:overuse} can be seen in Table~\ref{tab:overview}, Figure~\ref{fig:bar} and Table~\ref{tab:reg_AI}. Table~\ref{tab:overview} highlights the substantial effect of AI on reducing overuse across different incentive scheme treatments. Notably, the transition from pre-AI to post-AI conditions shows marked improvements: in the \Flat{} treatment, overuse decreased from 24.3\% to 16.0\%, a deduction rate of 34\%. Similar positive shifts are observed in \Progressive{} and \Regressive{} treatments, with significant reduction rates of 26\% and 62\% in overuse, respectively, all supported by statistically significant $p$-values.%
\footnote{Across all treatments, the rate of overuse is decreased from 36.1\% to 22.9\% from pre- to post-AI decisions. The overall deduction rate is at 36.7\%.}

Figure~\ref{fig:bar_accurate} visually complements these findings by illustrating the deduction of overuse rate in post-AI conditions. The left figure depicts the average instances of overuse times per 20 rounds, and the right figure further decomposes the average overuse proportion among participants by round number. Overall, we see a consistent deduction in overuse after the integration of AI in decision-making throughout the whole experiment. 

Finally, Table~\ref{tab:reg_AI} provides a detailed regression analysis showing the statistical impact of AI on medical overuse. The coefficients for PostAI across various models consistently indicate significant decreases in overtreatment. Model 5 further reveals the nuanced interaction between incentive schemes and the effect of AI; it can be seen from the table that the deduction effect of AI is almost twice as much as in \Regressive{} compared to the \Flat{} treatment ($p<0.1$). The effects remained robust after controlling for individual characteristics and professional ability.%
\footnote{Recall that professional ability or innate ability is measured by the accuracy rate in the 10 practice questions at the beginning of the experiment.}

\begin{figure}[h!]
    \centering
    \includegraphics[width=1\textwidth]{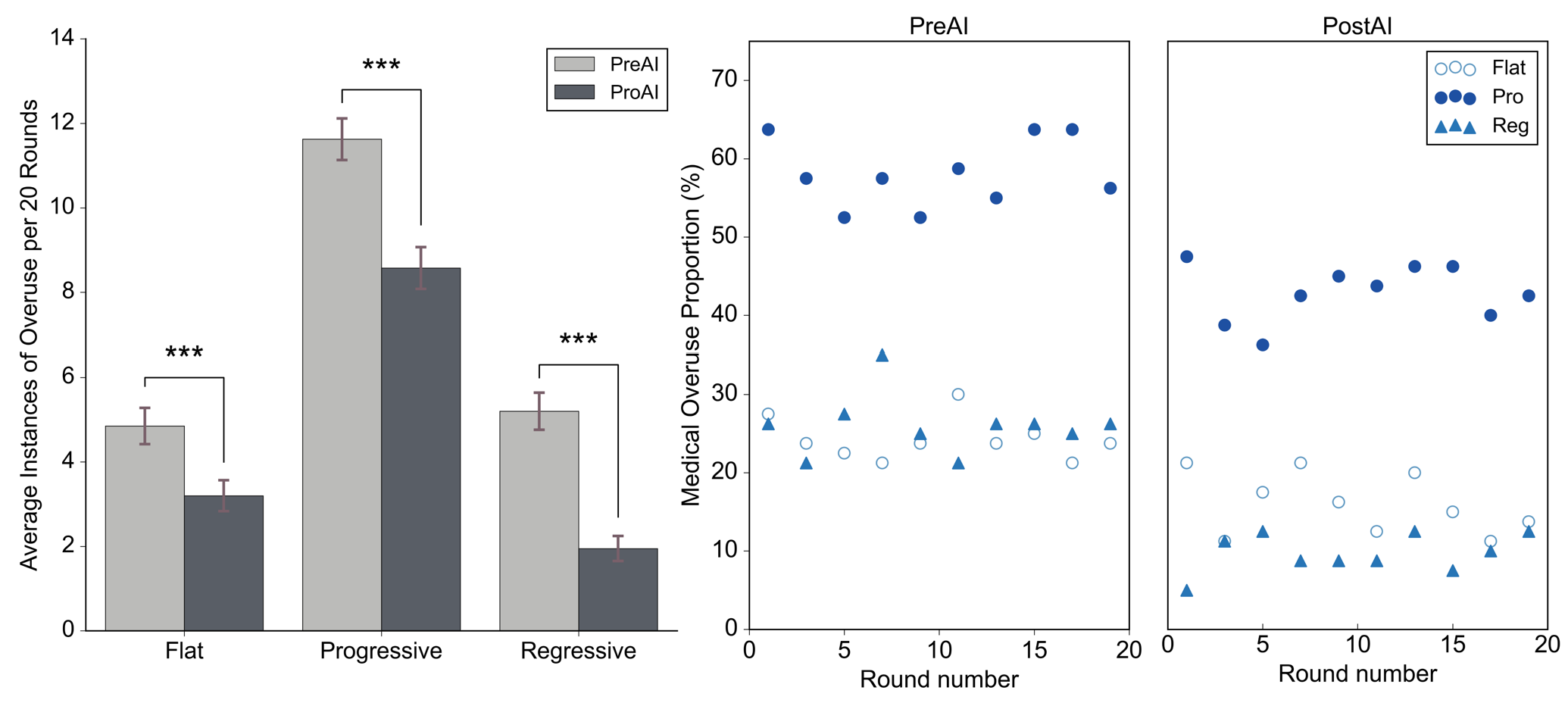} 
    \caption{Medical overuse by treatment}
    \label{fig:bar}
\end{figure}

\begin{table}[h!]\centering
\def\sym#1{\ifmmode^{#1}\else\(^{#1}\)\fi}
\caption{The effect of AI on frequency of medical overuse: Random effect models}
\begin{threeparttable}
\begin{tabular}{l*{5}{c}}
\toprule
                &\multicolumn{1}{c}{(1)}&\multicolumn{1}{c}{(2)}&\multicolumn{1}{c}{(3)}&\multicolumn{1}{c}{(4)}&\multicolumn{1}{c}{(5)}\\
\emph{Dep Var: Overuse freq }                &\multicolumn{1}{c}{\Flat}&\multicolumn{1}{c}{\Progressive}&\multicolumn{1}{c}{\Regressive}&\multicolumn{1}{c}{Pooled}&\multicolumn{1}{c}{Pooled}\\
\midrule
PostAI          &   -1.650***&   -3.050***&   -3.250***&   -2.650***&   -1.650***\\
(\textit{Baseline: PreAI})                &  (0.514)   &  (0.921)   &  (0.651)   &  (0.413)   &  (0.532)   \\
[1em]

Progressive     &            &            &             &            &    6.818***\\
(\textit{Baseline: Flat})                &            &            &            &            &  (1.229)   \\
[1em]
Regressive      &            &            &            &            &   -1.008   \\
                &            &            &            &            &  (1.525)   \\
[1em]

Progressive $\times$ PostAI&            &            &            &            &   -1.400   \\
                &            &            &            &            &  (1.091)   \\
[1em]

Regressive $\times$ PostAI&            &            &            &            &   -1.600*  \\
                &            &            &            &            &  (0.858)   \\
[1em]
Ability         &            &            &            &            &   -0.840***\\
                &            &            &            &            &  (0.269)   \\
[1em]
\multicolumn{5}{l}{Control for individual characteristics} & \checkmark \\
Constant     &    4.850***&    11.630***&    5.200***&    7.225***&    26.180***\\
                &  (0.762)   &  (1.049)   &  (0.782)   &  (0.575)   &  (7.545)   \\

\midrule
\textit{Mean of Y} &4.025& 10.100&3.575 &5.900 &5.900\\
Observations    &       80   &       80   &       80   &      240   &      240   \\
$R^2$           &   0.033   &   0.046  &    0.138   &    0.045   &    0.372   \\
\bottomrule
\end{tabular}
\begin{tablenotes}
    \small 
    \item{\textit{Notes. }This table reports the effect of incentive schemes and AI assistance on medical overuse using random effect models. Unit of observation is individual \emph{frequency of medical overuse out of 20 rounds}, pre- and post-AI.  Model 5 additionally controls for other individual characteristics, including gender, age, major of study, cognitive reflection, cognitive ability, risk preference, altruism preference, trust, algorithm literacy, trust in algorithms, awareness of algorithms, and perceptions of algorithm fairness. All standard errors are clustered at the subject level and are reported in parentheses. In this and the following table $* p<0.10, ** p<0.05, *** p<0.010$. }
\end{tablenotes}
\end{threeparttable}
\label{tab:reg_AI}
\end{table}

\noindent\textbf{Causes of overuse. }Recall that the causes of medical overuse are classified into two categories: monetary and non-monetary. Using the \Flat{} and \Regressive{} treatment, we established a baseline measure of the non-monetary incentives of overtreatment, as in these two treatments, overuse is associated with either no extra or even negative rewards. Indeed, we see a similar rate of overuse in these two treatments before the implementation of AI; the additional overuse in the \Progressive{} treatment shall then be attributed to monetary incentives. Using back-in-the-envelop calculations, we estimate the non-monetary incentive accounts for around 43\% of medical overuse, while the monetary incentive accounts for 57\%.
    
\end{proof}

\subsection{The effect of AI on healthcare quality}

\begin{result}
    The use of AI significantly increased the accuracy rate in the medical prescription task, with improvements ranging from 17\% to 37\%. The effect of AI on minimising unnecessary incorrect choices is strongest in the \Regressive{} treatment. 
\label{res:accuracy}
\end{result}

\begin{proof}[Support. ]
Support for Result~\ref{res:accuracy} is provided by Table~\ref{tab:overview}, Table~\ref{tab:reg_accuracy} and Figure~\ref{fig:bar_accurate}. Table~\ref{tab:overview} details the improvement in accuracy rates across treatments. In the \Flat{} treatment, the accuracy rate increased from 56.3\% pre-AI to 77.3\% post-AI, demonstrating a substantial 37\% enhancement in decision-making quality. Similarly, in the \Progressive{} and \Regressive{} treatments, the accuracy rates increased by 23\% and 17\%, respectively, after AI assistance.%
\footnote{Across all three treatments, the accuracy rate of the medical prescription task increased from 12.6\% to 15.8\% after the introduction of AI advice. The overall accuracy enhancement effect of AI is thus at 25\%.}

Table~\ref{tab:reg_accuracy} provides a detailed regression analysis showing the statistical impact of AI on accuracy. The coefficients for PostAI across various models consistently indicate significant improvements in accuracy. Model 4 further reveals the nuanced interaction between incentive schemes and the effect of AI; it can be seen that the accuracy improvement effect of AI is more pronounced in the \Flat{} treatment compared to the \Regressive{} treatment ($p < 0.1$). The effects remain robust after controlling for individual characteristics and professional ability (Model 5).

\begin{table}[h!]\centering
\def\sym#1{\ifmmode^{#1}\else\(^{#1}\)\fi}
\caption{The effect of AI on accuracy frequency: Random effect models}
\begin{threeparttable}
\begin{tabular}{l*{5}{c}}
\toprule
                &\multicolumn{1}{c}{(1)}&\multicolumn{1}{c}{(2)}&\multicolumn{1}{c}{(3)}&\multicolumn{1}{c}{(4)}&\multicolumn{1}{c}{(5)}\\
                \emph{Dep Var: Accuracy freq }     &\multicolumn{1}{c}{\Flat}&\multicolumn{1}{c}{\Progressive}&\multicolumn{1}{c}{\Regressive}&\multicolumn{1}{c}{Pooled}&\multicolumn{1}{c}{Pooled}\\
\hline
PostAI          &    4.200***&    3.050***&    2.275***&    4.200***&    4.200***\\
(\textit{Baseline: PreAI})   &  (0.535)   &  (0.546)   &  (0.501)   &  (0.532)   &  (0.553)   \\
[1em]

Progressive     &            &            &            &    2.200** &    1.514** \\
(\textit{Baseline: Flat}) &            &            &            &  (0.868)   &  (0.601)   \\
[1em]
Regressive      &            &            &            &    1.975** &    0.924   \\
                &            &            &            &  (0.818)   &  (0.635)   \\
[1em]

Progressive $\times$ PostAI&            &            &            &   -1.150   &   -1.150   \\
                &            &            &            &  (0.761)   &  (0.791)   \\
[1em]

Regressive $\times$ PostAI&            &            &            &   -1.925***&   -1.925** \\
                &            &            &            &  (0.729)   &  (0.757)   \\
[1em]
Ability         &            &            &            &            &    0.239** \\
                &            &            &            &            &  (0.105)   \\
[1em]
\multicolumn{5}{l}{Control for individual characteristics} & \checkmark \\

Constant        &    11.250***&    13.450***&    13.225***&    11.250***&    12.985***\\
                &  (0.611)   &  (0.621)   &  (0.549)   &  (0.609)   &  (1.981)   \\
\midrule
\textit{Mean of Y} &13.350 & 14.975&14.363 &14.229 &14.229\\
Observations    &       80   &       80   &       80   &      240   &      240   \\
$R^2$           &    0.361   &    0.219   &    0.167   &    0.294   &    0.569   \\
\bottomrule

\end{tabular}
\begin{tablenotes}
    \small 
    \item{\textit{Notes. }This table reports the effect of incentive schemes and AI assistance on accuracy frequency using random effect models. Unit of observation is individual \emph{frequency of being accurate out of 20 rounds}, pre- and post-AI.  Model 5 additionally controls for other individual characteristics, including gender, age, major of study, cognitive reflection, cognitive ability, risk preference, altruism preference, trust, algorithm literacy, trust in algorithms, awareness of algorithms, and perceptions of algorithm fairness. All standard errors are clustered at the subject level and are reported in parentheses. }
\end{tablenotes}
\end{threeparttable}
\label{tab:reg_accuracy}
\end{table}

\noindent\textbf{Mechanisms of Improvement. }Finally, Figure~\ref{fig:bar_accurate} visually dissects two potential channels of accuracy improvement, elucidating the mechanisms through which AI enhances decision-making:
\begin{enumerate}
    \item \textbf{Increase of one exact answer (\cmark)} - This improvement reflects the adoption of what we term the \textit{Optimal Treatment Adjustment}, where participants aim for the most exact and accurate strategy.
    \item \textbf{Increase of one correct plus one incorrect answer (\cmark\xmark)} - This pattern suggests a \textit{Precision-Compromise Strategy}, where participants make conservative choices, including a mix of correct and incorrect options.
\end{enumerate}

\begin{figure}[h!]
    \centering
    \includegraphics[width=0.7\textwidth]{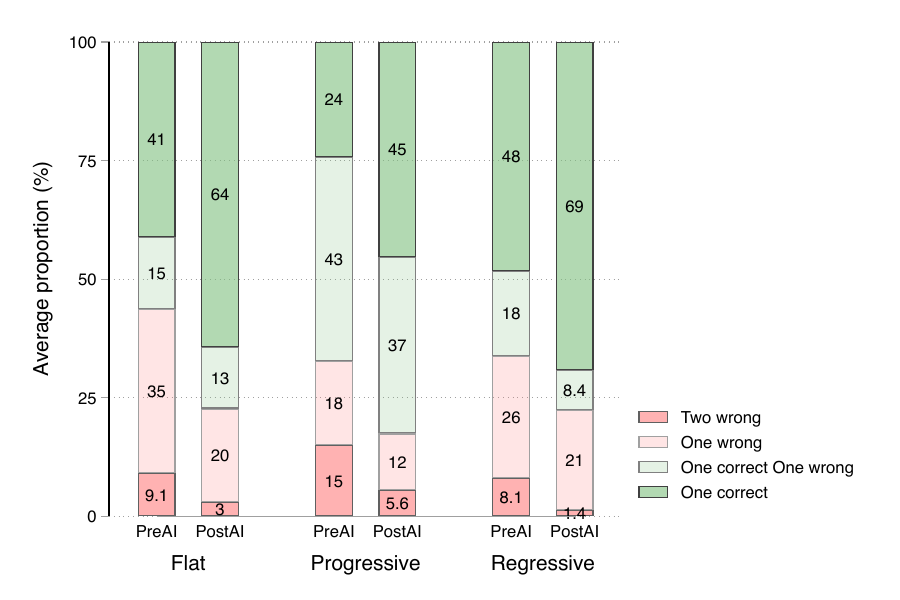}
    \caption{Accuracy rate in the medical prescription task}
    \label{fig:bar_accurate}
\end{figure}

The figure highlights that the relative significance of these enhancement channels varies across treatments. 
We quantify the relative contribution of the two accuracy enhancement mechanisms in each treatment as follows. In the \Flat{} treatment, the Optimal Treatment Adjustment (Channel 1) accounts for 109.5\% of the total enhancement effects, while the Precision-Compromise Strategy (Channel 2) detracts by -9.5\%. In the \Progressive{} treatment, Channel 1 contributes 140\%, and Channel 2 reduces the enhancement by -40\%. Finally, in the \Progressive{} treatment, Channel 1 leads to an enhancement of 184\%, with Channel 2 diminishing the effect by -84\%. 

Overall, AI predominantly enhances decision-making through Optimal Treatment Adjustment, slightly offset by a crowing-out effect on the Precision-Compromise Strategy. The influence of AI in reducing the unnecessary incorrect choice is the most pronounced in the \Regressive{} treatment.

\end{proof}

\subsection{Incentive scheme and social welfare}

\begin{result}[Incentive Schemes and Accuracy]
\label{res:incentives}
    Compared to \Flat{}, a \Progressive{} or a \Regressive{} incentive scheme significantly increases the initial accuracy rate before participants receive AI advice. However, the influence of the incentive schemes on performance becomes less pronounced once AI advice is introduced.
\end{result}

\begin{proof}[Support. ]
Support for this result comes from Table~\ref{tab:overview}, Table~\ref{tab:reg_accuracy} and Table~\ref{tab:welfare} (Panel B). Prior studies often advocate that performance-based pay can enhance physician performance more effectively than service-based pay schemes \citep{heider2020effects}. Our results in the PreAI conditions can test this hypothesis in the context of the medical prescription task context. As shown in Table~\ref{tab:overview}, in the \NoAI{} conditions, the accuracy rate is 56.3\% in \Flat{} which is lower than in \Progressive{} (67.3\%) and \Regressive{} (66.1\%). Table~\ref{tab:reg_accuracy} (Model 4) complements this comparison by showing that in PreAI conditions, the accuracy rate in \Flat{} treatment is significantly lower compared with the \Progressive{} and the \Regressive{} treatment. 
These comparisons remain robust in \Progressive{} treatment after controlling for individual characteristics and ability, but not for \Regressive{} treatment. These results suggest the benefit of medical overuse: i.e., the more conservative choices (\cmark\xmark) under uncertainty could essentially increase the overall accuracy.

To delve deeper, in Table~\ref{tab:welfare} panel B, we break down participants' choice types by treatment. Compared with the baseline (\Flat{}), the pay-by-performance incentive scheme where participants got punished with incorrect answers (\Regressive{}) significantly increased the exactly correct choice (\cmark) from 41\% to 48\%. Conversely, a pay-by-service incentive scheme where participants were rewarded with the number of medicines prescribed regardless of accuracy (\Progressive{}) significantly boosted the accuracy rate through conservative choices (\cmark\xmark).

Post-AI introduction, the effects of these incentive effects are mitigated, as seen from the significant negative interaction terms in the \Regressive{} treatment in Table~\ref{tab:reg_accuracy}, indicating that the quality-enhancing effects of a performance-contingent incentive scheme are equated after the introduction of AI. In other words, the effect of AI on enhancing healthcare quality is as effective as introducing a performance-based incentive scheme.%
\footnote{Additional evidences can be seen from Table~\ref{tab:overview}. In \NoAI{} conditions, the accuracy rate is significantly lower in \Flat{} than the \Regressive{} treatment ($p=0.014$). \AI{}, the accuracy rate disparity disappeared ($p=0.969$, \Flat{} vs \Regressive{}, \AI{}).}

\textbf{Discussion:} The literature consistently highlights that while performance-based incentives can motivate doctors to improve care quality, they must be carefully designed to avoid unintended consequences such as overtreatment or neglect of non-incentivized activities \citep{einav2018provider,petersen2006does,mcdonald2007impact,kovacs2020pay,fainman2020design}.%
\footnote{See \cite{van2010systematic} for a systematic review.}
From a practical perspective, even if altering medical incentive schemes is effective at promoting patient care quality, it may be difficult to implement in hospitals due to system inertia \citep{agrawal2022power}. Our results indicate that the introduction of AI appears to level the playing field among different incentive schemes in terms of their influence on physician performance (accuracy rate), simplifying incentive structures without compromising care quality.

\end{proof}

\begin{result}[Incentive Schemes and Social Welfare]
    Patients' payoffs are highest under the \Regressive{} treatment, followed by \Flat{} and \Progressive{} treatments: $\Regressive > \Flat = \Progressive$. For physicians, the \Regressive{} treatment consistently yields the highest payoffs, with the other two treatments being equal to each other: $\Regressive>\Progressive=\Flat$. Social welfare is maximized under the \Regressive{} treatment, which is particularly significant when considering different scenarios of benefits and costs of correct/incorrect treatments.
\end{result}

\begin{proof}[Support.]
    Support for the result can be seen in Table~\ref{tab:welfare}. Previous results have shown the accuracy enhancement and overuse reduction effects of AI assistance; we, therefore, would reasonably expect welfare enhancement after AI, conditional on the AI advice being costless. In this section, we will quantify the welfare enhancement of AI under each incentive scheme.

    On the other hand, the effect of incentive schemes on welfare is nuanced. Since overtreatment could potentially increase the accuracy rate, it is uncertain that a \Progressive{} incentive schemes that promote overtreatment would make the patients better or worse off overall. We, therefore, compare the effects of incentive schemes on patients' welfare before and after AI implementation with different costs of overtreatment.

    Table~\ref{tab:welfare} breaks down the welfare analysis under different incentives by various combinations of benefit ($B$) and costs ($C$) of correct(incorrect) treatment. Although we implemented the first scenario in our experiment, the pattern of the flat, progressive and regressive incentive schemes remains consistent across different scenarios.%
    \footnote{We derive the payoffs in the other scenarios in the following method: We first compute the hypothetical patients' payoffs based on the combination of the corresponding combination of $B$ \& $C$. Then, we adjust the physicians' payoffs in the \Regressive{} treatment using the same ratio as the patients' payoffs (patients:physicians = 1.5:1). Next, we adjust the physicians' payoffs in the \Flat{} and \Progressive{} treatments, ensuring that the average payoffs across the three incentive schemes remain constant. Detailed computations are reported in Table~\ref{tab:breakdown} in the Appendix.}

\begin{table}[h!]
    \centering
    \caption{Welfare analysis and Physicians' choices}
    \begin{threeparttable}
        \setlength{\tabcolsep}{3mm}{
            \begin{tabular}{ccccccc}
                \toprule 
                & \multicolumn{3}{c}{PreAI} & \multicolumn{3}{c}{PostAI} \\ 
                \cmidrule(r){2-4} \cmidrule(r){5-7}
                & \Flat & \Progressive & \Regressive & \Flat & \Progressive & \Regressive \\
                \midrule
                \multicolumn{7}{l}{\textbf{Panel A: Welfare Analysis}} \\
                \midrule
                \textit{Patients' payoffs} \\
                B=4 C=2       & 4.89  & 4.87 & \textbf{5.45} & 6.32 & 6.09 & \textbf{6.46} \\
                B=2 C=4       & 2.41  & 1.71 & \textbf{2.93} & 4.00 & 3.24 & \textbf{4.26} \\
                B=4 C=4       & 3.53  & 3.06 & \textbf{4.25} & 5.54 & 4.89 & \textbf{5.81} \\
                B=4 C=0       & 6.25  & \textbf{6.69} & 6.65 & 7.09 & \textbf{7.30} & 7.10 \\
                \textit{Physicians' payoff} \\
                B=4 C=2       & 3.00  & 3.16 & \textbf{4.21} & 3.00 & 2.86 & \textbf{4.86} \\
                B=2 C=4       & 0.75  & 0.79 & \textbf{2.44} & 0.75 & 0.71 & \textbf{3.24} \\
                B=4 C=4       & 1.50  & 1.58 & \textbf{3.43} & 1.50 & 1.43 & \textbf{4.40} \\
                B=4 C=0       & 4.50  & 4.74 & \textbf{4.98} & 4.50 & 4.29 & \textbf{5.33} \\
                \textit{Social welfare} \\
                B=4 C=2       & 7.89  & 8.03 & \textbf{9.66} & 9.32 & 8.95 & \textbf{11.32} \\
                B=2 C=4       & 3.16  & 2.50 & \textbf{5.37} & 4.75 & 3.95 & \textbf{7.50} \\
                B=4 C=4       & 5.03  & 4.64 & \textbf{7.68} & 7.04 & 6.32 & \textbf{10.21} \\
                B=4 C=0       & 10.75  & 11.43 &  \textbf{11.63} & 11.59 & 11.59 &  \textbf{12.43} \\
                \midrule
                \multicolumn{7}{l}{\textbf{Panel B: Choice Analysis}} \\
                \midrule
                \textit{Choice type} \\
                \multicolumn{1}{l}{Correct (\cmark)}               & 41.1\% & 24.3\% & 48.3\% & 64.3\% & 45.3\% & 69.1\% \\
                \multicolumn{1}{l}{Correct/Incorrect (\cmark\xmark)} & 15.1\% & 43.0\% & 17.9\% & 13.0\% & 37.3\% & 8.4\% \\
                \multicolumn{1}{l}{Incorrect (\xmark)}             & 34.6\% & 17.6\% & 25.8\% & 19.8\% & 11.8\% & 21.1\% \\
                \multicolumn{1}{l}{Double Incorrect (\xmark\xmark)} & 9.1\%  & 15.1\% & 8.1\%  & 3.0\%  & 0.5\%  & 1.4\% \\
                \bottomrule 
            \end{tabular}
        }
        \begin{tablenotes}
            \item{\small \textit{Notes:} This table reports (a) Welfare analysis centred around physicians' payoff, patients' benefits, and societal welfare, and (b) Physicians' choice statistics with respect to incentive scheme treatments (\Flat, \Progressive, \Regressive) before and after AI advice.}
        \end{tablenotes}
    \end{threeparttable}
    \label{tab:welfare}
\end{table}

    \begin{enumerate}
        \item \textbf{Scenario 1 ($B=4, C=2$): }In this scenario, the benefit of correct treatment is high, and the cost of incorrect treatment is moderate. The \Regressive{} incentive scheme, characterized by decreasing rewards for each additional incorrect diagnosis, strikes an optimal balance by encouraging optimal treatment (\cmark) rather than excessive conservative adjustments (\cmark\xmark). Despite a higher accuracy rate in \Progressive{} treatment due to an increase in the precision-compromise strategy, patients' welfare is not significantly improved in \Progressive{} due to the moderate cost of overtreatment. 
        \item \textbf{Scenario 2 ($B=2, C=4$): }Here, the cost of incorrect treatment is twice as high as the benefit of correct treatment, placing a more significant penalty on overtreatment. In this case, the advantageous position of \Regressive{} compared to \Progressive{} becomes more pronounced in terms of patients' payoffs. As the cost of incorrect treatment increases, the benefit of \Progressive{} treatment encouraging optimal treatment adjustments becomes more appealing.
        \item \textbf{Scenario 3 ($B=4, C=4$): }In this scenario, the benefit of correct treatment equals the cost of incorrect treatment. Similar to the previous scenarios, the \Progressive{} scheme maximizes both physicians' and patients' payoffs. The differences are augmented after AI assistance.
        \item \textbf{Scenario 4 ($B=4, C=0$): }Here, we assume there is no cost for incorrect medicine, making overtreatment harmless. The \Progressive{} schemes encouraging more medical prescriptions appeared to be optimal for the patients' welfare. However, physician's payoffs are still maximized under the \Regressive{} scheme due to the largest proportion of optimal treatment (\cmark), and the total social welfare is still maximized in the \Regressive{} treatment.
    \end{enumerate}

Taken together, the \Regressive{} treatment maximizes patients' payoffs (except when $C=0$), physicians' payoffs, and social welfare, particularly when augmented by AI assistance. The robust performance of the \Regressive{} scheme across different scenarios of $B$ and $C$ values suggests that it may be a preferred approach in various contexts, offering a balance that optimizes welfare while accounting for the complexities of medical decision-making.
    
\end{proof}

\subsection{Further analysis: AI adoption and individual heterogeneity}

\begin{result}[Determinants of AI-adoption]
    Participants with higher scores in algorithm trust are more likely to adopt AI advice. Conversely, participants with higher confidence, higher ability, or those more senior in medical school are less likely to follow AI advice, ceteris paribus.
\end{result}

\begin{proof}[Support.]

Support for this result is provided by Table~\ref{tab:reg_adopt}.Previous results demonstrated the substantial effect of AI on reducing overuse and improving accuracy. Here, we delve into the behavioral question of who is more likely to adopt AI advice in their decision-making. We explore treatment differences and individual heterogeneities in AI adoption.

Table~\ref{tab:reg_adopt} shows that AI adoption is marginally higher in the \Progressive{} treatment using the main measures due to higher levels of affirmations (e.g., initial choice-AI advice-second choice follows A-A-A). However, the treatment differences disappear when we use the adjusted measure of AI adoption, which omits the affirmation of the adoption of AI.

\begin{table}[h!]\centering \small
\def\sym#1{\ifmmode^{#1}\else\(^{#1}\)\fi}
\caption{Determinants of AI adoption: Random effect models}
\begin{threeparttable}
\setlength{\tabcolsep}{1mm}{
\begin{tabular}{l*{6}{c}}
\toprule
               &\multicolumn{1}{c}{(1)}&\multicolumn{1}{c}{(2)}&\multicolumn{1}{c}{(3)}&\multicolumn{1}{c}{(4)}&\multicolumn{1}{c}{(5)}&\multicolumn{1}{c}{(6)}\\
Dep Var: AI adoption    &\multicolumn{1}{c}{Main}&\multicolumn{1}{c}{Main}&\multicolumn{1}{c}{Main}&\multicolumn{1}{c}{Adjust}&\multicolumn{1}{c}{Adjust}&\multicolumn{1}{c}{Adjust}\\
\midrule

\Progressive     &   0.024*  &   0.026** &   0.032***&  -0.012   &   0.029   &   0.046   \\
(\textit{Baseline: \Flat})  & (0.012)   & (0.012)   & (0.011)   & (0.052)   & (0.043)   & (0.039)   \\
[1em]
\Regressive      &   0.015   &   0.016   &   0.012   & -0.008   &   0.017   &  -0.019   \\
                & (0.012)   & (0.012)   & (0.014)   & (0.044)   & (0.037)   & (0.051)   \\
[1em]
Ability         &            &  -0.034   &  -0.031   &            &   -0.643***&   -0.469***\\
                &            & (0.024)   & (0.032)   &            & (0.071)   & (0.094)   \\
[1em]
Belief          &            &            &  -0.036   &            &            &   -0.229** \\
                &            &            & (0.035)   &            &            &  (0.109)   \\
[1em]
Surgery         &            &            &  -0.036** &            &            &  -0.083*  \\
(\textit{Baseline: Internal Medicine}) &            &            & (0.015)   &            &            & (0.044)   \\
[1em]
Grade           &            &            & -0.005   &            &            &  -0.062***\\
                &            &            &(0.005)   &            &            & (0.017)   \\
[1em]
AI trust        &            &            &  0.005   &            &            &   0.086** \\
                &            &            & (0.011)   &            &            & (0.035)   \\
[1em]
\multicolumn{3}{l}{Control for other individual characteristics} &  \checkmark & & & \checkmark \\ \\
Constant        &    0.952***&    0.969***&    0.990***&    0.489***&    0.810***&    0.827***\\
                & (0.010)   & (0.016)   & (0.074)   & (0.034)   & (0.043)   &  (0.227)   \\
\hline
Observations    &     2400   &     2400   &     2400   &     2400   &     2400   &     2400   \\
\bottomrule
\end{tabular} }
\begin{tablenotes}
    \small \item{\textit{Notes. }This table reports the factors that influence AI adoption rate using random effect models. The unit of observation is at the individual-round level, totalling 120 subjects $\times$ 20 rounds of observations. Robust standard errors are clustered at individual levels and are reported in parentheses. Model 3 \& 5 control for other individual characteristics, including gender, age, risk attitudes, altruistic preferences, trust, cognitive ability (measured by the CRT \& Raven tests, and other algorithm preferences (awareness, literacy and fairness perceptions). }
\end{tablenotes}
\end{threeparttable}
\label{tab:reg_adopt}
\end{table}

We observe notable individual heterogeneity in the tendency to adopt AI advice:%
\footnote{We also have additional results regarding individual heterogeneity on the effect of AI on medical overuse propensity and accuracy rate, considering space and relevance, these results are reported in the Appendix (Result~\ref{res:indv_hetero}).}

\begin{itemize}
    \item \textbf{Ability}: Measured by the accuracy rate in incentivized practice questions at the beginning of the experiment, ability is negatively correlated with adjusted measures of AI adoption (Models 5 and 6). This is consistent with existing literature indicating that experts exhibit higher algorithm aversion than laypeople \citep{kawaguchi2021will}.
    \item \textbf{Belief: }Measured by self-reported belief in the accuracy rate in practice questions, which can be seen as a measure of self-confidence, is also negatively related to AI adoption (Model 6). This aligns with the general finding that overconfidence is associated with greater algorithm aversion  \citep{burton2020systematic}.
    \item \textbf{AI trust: }Measured by self-reported trust in AI in everyday life using post-survey questionnaires, AI trust is positively correlated with AI adoption (Model 6), suggesting that our AI trust measures are good indicators of preferences towards AI advice.
    \item \textbf{Grade: }We found that higher-grade medical students are less likely to adopt AI advice, controlling for ability and self-reported beliefs in accuracy (Model 6).
\end{itemize}

Most interestingly, we found that students who majored in Surgery, compared with internal medicine, were significantly less likely to change their choices after seeing AI advice in both main and adjusted measures of AI adoption (Model 3 \& 5). We believe it is the first evidence of heterogeneity in algorithm appreciation propensity among medical specialities. Although these results should be further tested for their robustness and generalizability, we hypothesize that this may be due to the selection effect of the surgery speciality. Individuals who are more likely to trust their own intuition and practice rather than outsource information may be more inclined to choose a surgical discipline compared with Internal medicine.%
\footnote{For example, existing studies found that surgical residents tend to exhibit higher levels of assertiveness and decisiveness, traits associated with a lower likelihood of relying on external advice; In contrast, internal medicine residents, who typically engage in more collaborative and deliberative decision-making processes, may be more open to incorporating advice into their clinical judgments \citep{shanafelt2010burnout,hussenoeder2021comparing}. 
This evidence suggests that the disparity in AI advice adoption between surgery and internal medicine major students might stem from disparity in cognitive reasoning and judgements.
}

\end{proof}

\section{Discussions and Conclusion}
\label{sec:conclusion}

Our findings underscore that integrating AI significantly curbs medical overtreatment, achieving reductions of up to 62\% under aligned incentive structures (\Regressive{}). This phenomenon termed the \emph{Synergistic AI Effect}, highlights the critical importance of incentive alignment in maximizing AI's efficacy. In terms of quality, AI intervention not only reduced overtreatment but also enhanced diagnostic accuracy across all incentive schemes. The implementation of AI led to more precise medical prescriptions, particularly evident in the shift towards the ``Optimal Treatment Adjustment'' strategy in \Regressive{} treatment, which optimized the quality of care without increasing overtreatment.

The study identifies two pivotal causes of overtreatment: monetary incentives and non-monetary factors such as defensive medicine and knowledge gaps. Our results indicate that around 43\% of medical overuse is due to non-monetary incentives—participants exhibit overuse even if the extra choice is not rewarded (\Flat{}) or even negatively rewarded (\Regressive{}). Conversely, monetary incentives account for 57\% of total medical overuse in the \Progressive{} treatment, where overtreatment is financially incentivized.

Based on this dichotomy, we propose two potential interventions to curb medical overuse and improve healthcare quality: 1) Implement a pay-by-performance incentive scheme where overtreatment is penalized, such as the \Regressive{} incentive schemes in our experiment, to align the interests of physicians and patients;  2) Introduce human-AI collaboration by incorporation AI advice in decision making. 
From a policy perspective, we show that the effects of these two interventions at ensuring healthcare quality can be comparable within existing incentive structures environment, such as pay-by-visit (\Flat) and pay-by-service (\Progressive{}), pointing to a promising avenue for integrating AI into the healthcare system.

Theoretically, our findings can be interpreted through the lens of principal-agent theory, where AI acts as a mechanism to align the interests of healthcare providers (agents) with those of patients and payers (principals) by reducing non-monetary incentives of overtreatment. The effectiveness of AI in reducing overtreatment could be modelled as a function of its ability to provide transparent, unbiased, and evidence-based recommendations, which counteract the misaligned incentives inherent in many healthcare systems.

\noindent\textbf{Conclusion. }This study demonstrates how artificial intelligence (AI) can be pivotal in reducing medical overtreatment and enhancing diagnostic precision. Reflecting the quote at the beginning of this paper, Taleb advocates querying a doctor not just on general advice but on personal choices if placed in similar circumstances. Our research supports this through the results observed in the \Regressive{} treatment, where AI's impact was most profound when physician and patient interests were closely aligned -- seeking optimal exact treatment, thus maximizing welfare.

Drawing from ``Power and Prediction'', the introduction of AI in healthcare signifies a shift from conservative protocols towards a system driven by predictive analytics and tailored treatments \citep{agrawal2022power}. This evolution in healthcare suggests that AI's role can extend beyond mere assistance to being a fundamental part of strategic health management, focusing on prevention and precision. Our study takes the first step in mimicking the results and consequences of AI-physician synergy in a harmless, risk-free environment.

Future research should extend these findings by exploring the long-term impact of AI across various healthcare systems while rigorously evaluating ethical dimensions to ensure that the integration of AI upholds principles of fairness and equity in patient care globally.


\bibliographystyle{johd}
{
\onehalfspacing \small
\bibliography{bib}
}

\clearpage

\appendix
\begin{appendices}

\setcounter{table}{0}
\setcounter{figure}{0}
\renewcommand{\thetable}{A\arabic{table}}
\renewcommand{\thefigure}{A\arabic{figure}}

\section{Additional tables and figure}
\subsection{Balance test across treatments}

We first report the balance tests across treatment groups among demographic variables and responses from the post-experimental questionnaires. Table~\ref{tab:BalanceTest} reports the mean and $p-values$ from Fisher's exact tests, examining the equality of distributions across three treatment groups. Despite a marginally higher (less than one year difference) age and grade in \Regressive{} treatment and a lower CRT scale, comparisons in all other categories are not statistically significant. The overall test of significance across all variables yields $F=6.12$.

\begin{table}[H]
    \centering
    \caption{Randomization balance checks}
    \begin{threeparttable}
    \begin{tabular}{lccccccc}
    \toprule
    \textbf{Individual characteristics} &\textbf{Flat} &\textbf{Prog}&\textbf{Reg }  & \textbf{Total} &\textbf{$p$-value} \\
    \midrule 
    Ability  & 5.00 & 5.70 & 5.38  &5.36 & 0.743\\
    Female(\%)& 67.50 &52.50  &57.50  &59.17 &0.430\\
    Age      &20.68  &20.80  & 21.30  &20.93 &0.000\\
    Grade    &3.35  & 3.58  &4.40   &3.78 &0.000\\
    Altruism &2.66  &2.484  &2.55  & 2.57& 0.312\\
    Trust    &3.55 &3.68  &3.98     &3.73 & 0.413\\
    CRT      &3.58 & 3.55 &2.03    &3.05 &0.000\\
    Raven    &5.15 &5.45  &5.5     &5.367 &0.017\\
    Risk     &3.20 & 3.16 &3.26   &3.21  & 0.962\\
    AI Trust &3.77 &3.65  & 3.76  & 3.72 & 0.175\\
    Fairness & 3.15  &3.03  &3.21 &3.13  & 0.657\\
    Literacy &3.48  & 3.50  & 3.5 & 3.49 &0.997\\
    Awareness& 3.97 & 3.96  &3.98 &3.97  &0.472\\
    Surgery(\%)&20.00  & 20.00  &15.00 &18.33 &0.347\\
    \bottomrule
    \end{tabular}
    \begin{tablenotes}
        \small
        \item \textit{Notes:}  This table reports the average values of individual characteristic variables across treatments. $p-$values are from the Fisher's exact test, treating each individual as an independent observation. With the exception of a few variables about Age, Grade, CRT, Raven, there were no significant differences in the distribution of individual characteristics between treatments. For an overall test of balance across the 14 variables,F-statistic=6.12.
        \end{tablenotes}
    \end{threeparttable}
    \label{tab:BalanceTest}
    \end{table}

\newpage
\subsection{AI adoption rate by treatment}

\begin{table}[h!]
    \centering
    \caption{AI-adoption Summary Statistics}
    \begin{threeparttable}
    \setlength{\tabcolsep}{10mm}{
    \begin{tabular}{lcc}
    \toprule
    &  Mean (\%) & SD \\
    \midrule
    \multicolumn{3}{l}{\textit{Panel A. Flat Treatment}} \\
    Main & 95.2 & 0.213   \\
    Adjusted & 48.9  & 0.500\\
    \midrule
    \multicolumn{3}{l}{\textit{Panel B. Progressive Treatment}} \\
    Main &  97.62 & 0.152 \\
    Adjusted & 47.3 & 0.500 \\
    \midrule
    \multicolumn{3}{l}{\textit{Panel C. Regressive Treatment}} \\
    Main & 96.8 & 0.177  \\
    Adjusted  & 48.1 & 0.500  \\
    \bottomrule
    \end{tabular}}
    \begin{tablenotes}
        \small
        \item \textit{Notes:} This table reports summary statistics for  AI adoption across treatments. The columns labeled ``Main'' and ``Adjusted'' refer to different AI adoption metrics defined in Table~\ref{tab:ai_adopt_def}.
    \end{tablenotes}
    \end{threeparttable}
    \label{tab:adoption_sum}
\end{table}
\clearpage

\newpage
\subsection{Individual heterogeneity in the effect of AI on medical overuse and accuracy}

\begin{result}[Individual heterogeneity]
\label{res:indv_hetero}
    The enhancement effect of AI on accuracy rate in the medical prescription task is more pronounced for junior than senior medical students; similarly, the enhancement effect of AI is more pronounced among low-ability than high-ability medical students.
\end{result}

\begin{proof}[Support. ]
Support for this result can be seen in Figure~\ref{fig:forestplot_accuracy}. 
Figure~\ref{fig:forestplot_accuracy} plots the coefficient and 95\% confidence interval of ``PostAI'' from random effect panel data models with participants' average accuracy rate as the dependent variable for each corresponding subsample: Confidence is measured by the difference in `belief' and `actual' performance in the 10 practice questions, with a higher belief than actual performance defined as overconfidence, and equal or lower belief defined as moderate (51\%); Medical students higher than the fourth year are defined as `Senior' (48\%), others are defined as `Junior'; Ability is measured by the accuracy rate in the practice questions, participants ranked in the upper half are defined as the high ability, the lower half is defined as low ability (50\%); Finally 59\% of the participants were female, the rest are male.

     \begin{figure}[h!]
        \centering
        \includegraphics[width=0.6\textwidth]{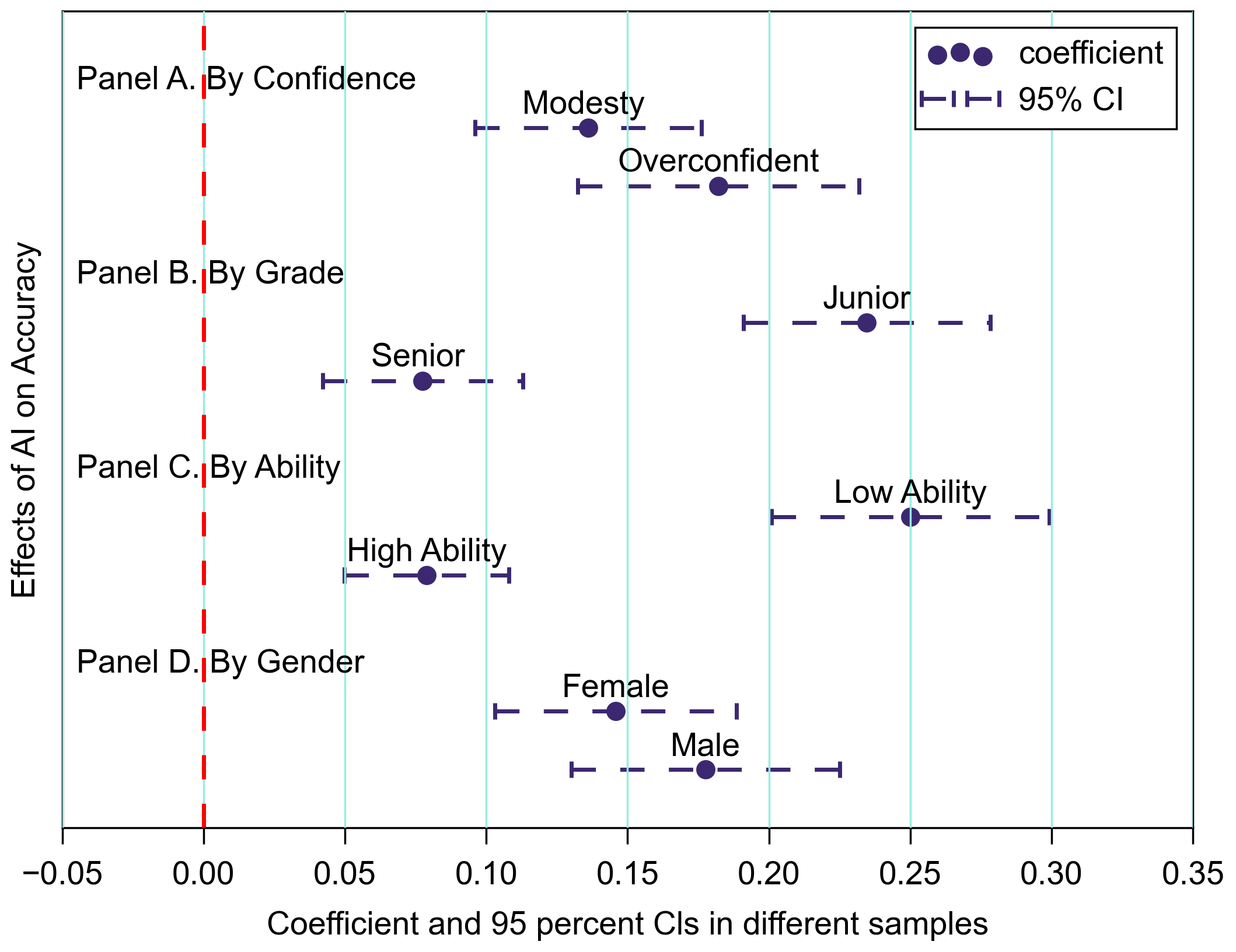}
        \caption{Individual heterogeneity in the effect of AI on accuracy}
        \label{fig:forestplot_accuracy}
    \end{figure}

It can be seen from the figure that, across all subgroups, the coefficient of PostAI is statistically significantly higher than 0, indicating a consistent positive effect of AI on the accuracy rate. Intriguingly, we found the enhancement effects are more pronounced for vulnerable groups who either have a lower ability based on the practice questions ($p<0.01$), or are at a lower grade (junior) at the medical school education ($p<0.01$).
This result aligns with another field study, which reported that the effect of ChatGPT on productivity improvement is most pronounced among low-skill workers \citep{li2024advanced}.

Notably, we also examined the individual heterogeneity on the effect of AI on overtreatment using a similar comparison but with individual average rates of overtreatment as the dependent variable (Figure~\ref{fig:forestplot_overuse}). However, we do not detect any significant disparity across any subgroup comparisons. We conclude that the effect of AI on overtreatment is consistent and homogeneous across all demographic groups.%
\footnote{AI trust in Figure~\ref{fig:forestplot_overuse} is measured by participants answers in the post-experimental questionnaires (\textit{Questionnaire 5: Algorithm fairness and trust}, available in the Appendix). We define participants who ranked in the upper half as `AI Trust', and the lower half as `No AI Trust'.
}

   \begin{figure}[h!]
        \centering
        \includegraphics[width=0.6\textwidth]{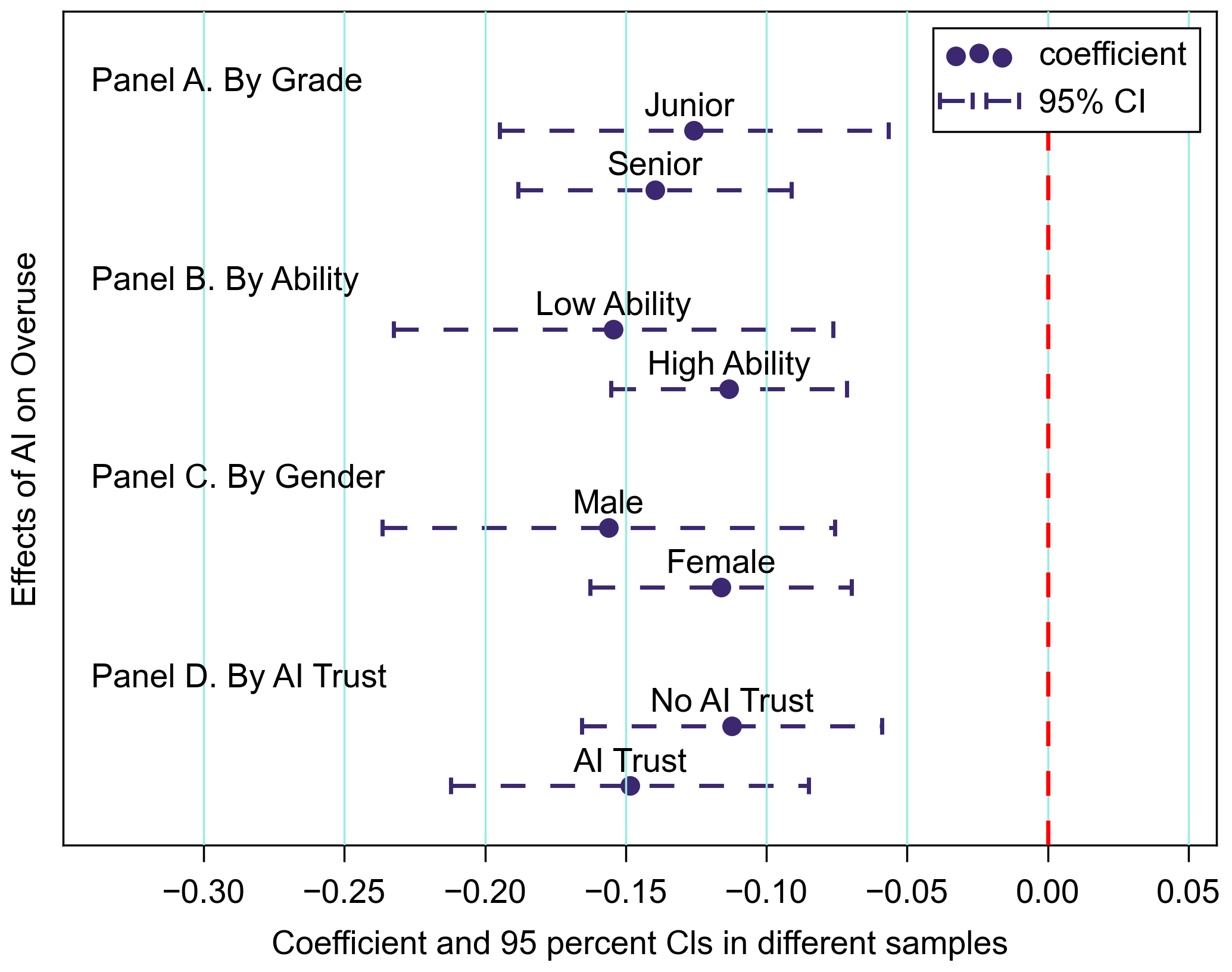}
        \caption{Individual heterogeneity in the effect of AI on overuse}
        \label{fig:forestplot_overuse}
    \end{figure}
\end{proof}

\newpage
\subsection{Additional tables for welfare analysis}
\begin{table}[H]
    \centering
    \caption{Calculation method of Physicians' and  Patients' payoff}
    \begin{threeparttable}
        \setlength{\tabcolsep}{3mm}{
        \begin{tabular}{lcccc}
        \toprule 
        & \multicolumn{3}{c}{Physicians' payoff} & Patients’ payoffs \\ 
        \cmidrule(r){2-4}
        & \Flat & \Progressive & \Regressive &   \\
        \midrule 
        \textit{B=4 C=2 } \\
        Single choice, Correct (\cmark) & 3 & 2 & 6 & 8 \\
        Double choice, Correct/Incorrect (\cmark\xmark) & 3 & 4 & 4.5 & 6\\
        Single choice, Incorrect (\xmark)  & 3 & 2 & 1.5 & 2 \\
        Double choice, Double Incorrect (\xmark\xmark)  & 3 & 4 & 0 & 0 \\
         \textit{B=2 C=4 } \\
        Single choice, Correct (\cmark) & 0.75  & 0.5  & 4.5 & 6 \\
        Double choice, Correct/Incorrect (\cmark\xmark) & 0.75 & 1 & 1.5 & 2  \\
        Single choice, Incorrect (\xmark) & 0.75 & 0.5 & 1.5 & 0\\
        Double choice, Double Incorrect (\xmark\xmark) & 0.75 & 1 & -3 & -4  \\
         \textit{B=4 C=4 } \\
        Single choice, Correct (\cmark) & 1.5 &1 & 6 & 8 \\
        Double choice, Correct/Incorrect (\cmark\xmark) &1.5 & 2 & 3 &4 \\
        Single choice, Incorrect (\xmark) &1.5 & 1 & 0 &0  \\
        Double choice, Double Incorrect (\xmark\xmark) & 1.5 & 2 &-3 & -4   \\
         \textit{B=4 C=0 } \\
        Single choice, Correct (\cmark) & 4.5 &3 &6 & 8 \\
        Double choice, Correct/Incorrect (\cmark\xmark) &4.5 & 6 &6 &8 \\
        Single choice, Incorrect (\xmark) &4.5 &3 &3 &4  \\
        Double choice, Double Incorrect (\xmark\xmark) &4.5 &6 &3 &4  \\
         \bottomrule 
        \end{tabular}}
        \begin{tablenotes}
            \item{\small \textit{Notes. }            
            This table outliens the derivation of physicians' and patients' payoffs under varying cost and benefit scenarios. The methodology is grounded in three guiding principles: (1) Patients' payoffs are directly derived from the respective $B$ and $C$ values, representing benefits and costs. (2) The physicians' payoff in the \Regressive{} treatment is calculated as a linear function of the corresponding patients' payoffs, maintaining a ratio of 3:4 (Physician to Patient). (3) In the \Flat{} and \Progressive{} treatments, the physicians' payoffs are structured to match the average payoff observed in the \Regressive{} treatment, adhering to incentive structures that are either constant or increasing with the number of medicine choices selected.}
            \end{tablenotes}
        \end{threeparttable}
        \label{tab:breakdown}
        \end{table}

\newpage
\section{Experimental Instructions}
\label{sec:instruction}

\begin{center}\Large\bfseries{Instructions}\end{center}

Welcome to the Wuhan University Medical Decision Making Experiment. Thank you for your support! This experiment simulates a real doctor-patient consultation scenario. The patient's basic information and disease introductions are selected from the ``National Qualifying Examination for Medical Practitioners" real question bank. Before the experiment begins, you will choose your preferred specialty or department based on your expertise or interest and then enter the diagnosis and treatment environment, waiting for the formal experiment to start.

This experiment consists of three parts:
\begin{itemize}
    \item Experimental Task 1: This part consists of 10 questions, with only 1 option per question.
    \item Experimental Task 2: This part consists of 20 questions where you can choose 1 or 2 options (multiple choice).
    \item Experimental Task 3: This part consists of 5 questionnaires.
\end{itemize}

\vspace{3mm}

\noindent
\textbf{Experimental Task 1: Single Choice Questions}

Task 1 includes a total of 10 single-selected questions, each with only \textbf{one} correct answer. After answering the 10 questions, you will be prompted to fill in the number of questions you believe you have answered correctly.

\begin{figure}[h!]
    \centering
    \includegraphics[width=0.8\textwidth]{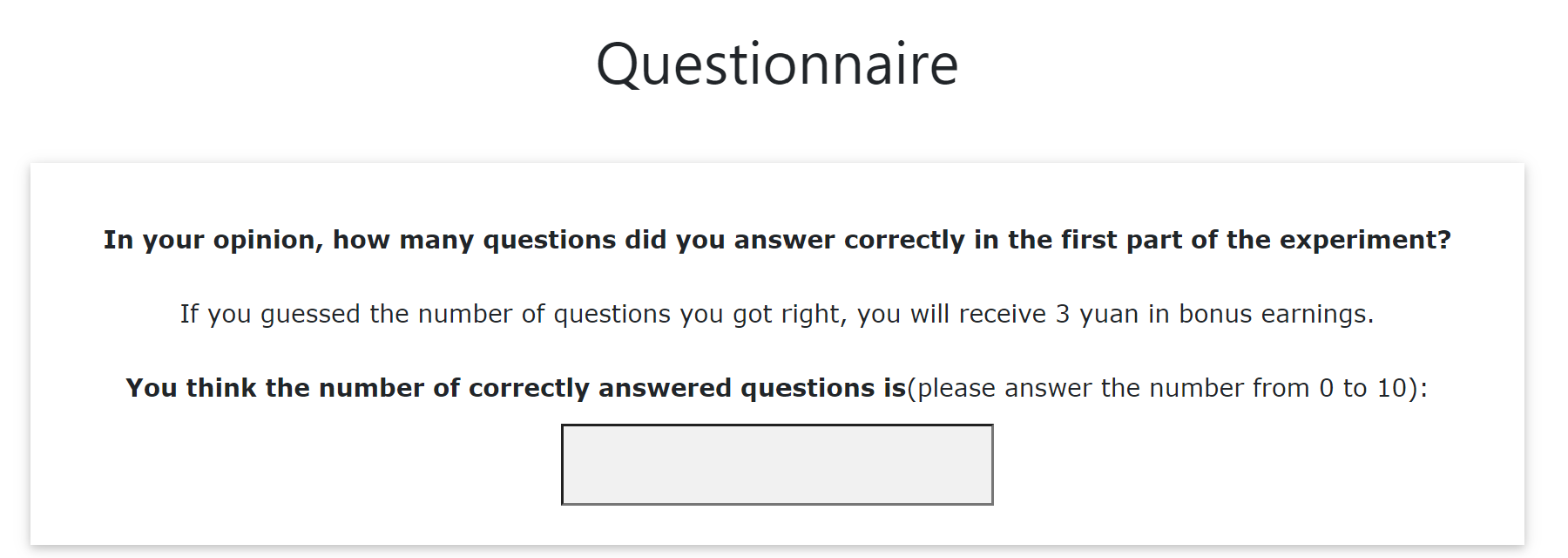}
    \end{figure}

\vspace{3mm}
\noindent
\textbf{Experimental Task 1 earnings}

For each correctly answered question out of the 10 single-selected questions, you will receive one yuan. Additionally, if your belief about the number of correct answers matches your actual performance, you will receive an extra 3 yuan reward.

\vspace{3mm}
 \begin{center}\Large\bfseries{Instructions (Experimental Task 2)}\end{center}
 \vspace{3mm}
 
\noindent
\textbf{Experimental Task 2: Medical Prescription Task}

The second part consists of 20 questions, each with only one correct answer, but you can choose between 1 or 2 options depending on the patient's condition. You will be given two opportunities to make a choice, labeled as the first choice and the second choice.
 \vspace{3mm}

\noindent\underline{Experimental steps }
\begin{itemize}
    \item First choice: Based on the patient’s basic conditions, you can choose1 option or 2 options.
    \item AI-Assisted Diagnosis and Treatment Suggestions: The system will provide you with artificial intelligence (AI) suggestions for diagnosis and treatment. The AI-assisted diagnosis and treatment suggestions introduced in this experiment are based on the ChatGPT-4.0 model developed by OpenAI. After fully studying the medical knowledge graph and database, we will use this question bank to The analysis suggestions given are not analysis of the answers to the real questions. Its average accuracy rate in different professional departments reaches 73.93\%.
    \item Second choice: You will make a second choice based on the AI suggestions, again choosing 1 option or 2 options.

\end{itemize}

After the experiment, we will announce whether your payment will be based on the first or second choice, so please take each choice seriously.
\\

\vspace{3mm}
\noindent
\textbf{Experimental Task 2 earnings}
\vspace{3mm}

\noindent{\underline{\textit{For FLat treatment}}}
\begin{itemize}
\item{\textbf{Doctors' Income:}} In this experiment, you will receive 3 yuan for each question completed, regardless of whether the answer is correct or incorrect.
\item{\textbf{Patients' Benefits:}} We will make a charitable donation to the Tencent Charity “Love Angel” program for children with leukemia and tumors based on the patient proceeds from your choices. If you choose one option and answer correctly, your patient will receive 8 yuan; if you answer incorrectly, your patient will receive 2 yuan; if you choose two options and answer correctly to one of them, your patient will receive 6 yuan; if you answer incorrectly to both options, your patient will receive 0 yuan.

\end{itemize}

\noindent{\textbf{The income matrix is as follows:}}

\begin{table}[H]
        \centering
        \begin{tabular}{lcc}                                                         \toprule
        Choice& Physicians’ payoff   & Patients’ payoffs    \\
        \hline
        Single choice ,Correct (\cmark)   & 3 yuan      & 8 yuan   \\
        Double choice,Correct/Incorrect (\cmark\xmark) & 3 yuan      & 6 yuan  \\
        Single choice ,Incorrect (\xmark)   & 3 yuan     & 2 yuan   \\
        Double choice,Double Incorrect (\xmark\xmark)  & 3 yuan     & 0 yuan  \\
        \bottomrule 
        \end{tabular}
        \end{table}

\noindent{\underline{\textit{For \Progressive{} treatment}}}
\begin{itemize}
\item{\textbf{Doctors' Income:}} In this experiment, you will receive 2 yuan for each option chosen, whether it is correct or incorrect.
\item{\textbf{ Patients' Benefits:}} We will make a charitable donation to the Tencent Charity “Love Angel” program for children with leukemia and tumors based on the patient proceeds from your choices. If you choose one option and answer correctly, your patient will receive 8 yuan; if you answer incorrectly, your patient will receive 2 yuan; if you choose two options and answer correctly to one of them, your patient will receive 6 yuan; if you answer incorrectly to both options, your patient will receive 0 yuan.

\end{itemize}

\noindent\textbf{The income matrix is as follows:}

\begin{table}[H]
        \centering
        \begin{tabular}{lcc}                                                         \toprule
        Choice& Physicians’ payoff   & Patients’ payoffs    \\
        \hline
        Single choice ,Correct (\cmark)   & 2 yuan      & 8 yuan   \\
        Double choice,Correct/Incorrect (\cmark\xmark) & 4 yuan      & 6 yuan  \\
        Single choice ,Incorrect (\xmark)   & 2 yuan     & 2 yuan   \\
        Double choice,Double Incorrect (\xmark\xmark)  & 4 yuan     & 0 yuan  \\
        \bottomrule 
        \end{tabular}
        \end{table}

\noindent\underline{\textit{For \Regressive{} treatment}}
\begin{itemize}
\item{\textbf{Doctors’ Income:}} In this experiment, if you choose one option and answer correctly, you will receive  6 yuan. If incorrect, you will receive 1.50 yuan. If you choose two options and one is correct, you will receive 4.50 yuan. If both are incorrect, you will receive 0 yuan.
\item{\textbf{Patients' Benefits:}} We will make a charitable donation to the Tencent Charity “Love Angel” program for children with leukemia and tumors based on the patient proceeds from your choices. If you choose one option and answer correctly, your patient will receive 8 yuan, if you answer incorrectly, your patient will receive 2 yuan; if you choose two options and answer correctly to one of them, your patient will receive 6 yuan, if you answer incorrectly to both options, your patient will receive 0 yuan.

\end{itemize}

\noindent\textbf{The income matrix is as follows:}

\begin{table}[H]
        \centering
        \begin{tabular}{lcc}                                                         \toprule
        Choice& Physicians’ payoff   & Patients’ payoffs    \\
        \hline
        Single choice ,Correct (\cmark)   & 6 yuan      & 8 yuan   \\
        Double choice,Correct/Incorrect (\cmark\xmark) & 4.5 yuan      & 6 yuan  \\
        Single choice ,Incorrect (\xmark)   & 1.5 yuan     & 2 yuan   \\
        Double choice,Double Incorrect (\xmark\xmark)  & 0 yuan     & 0 yuan  \\
        \bottomrule 
        \end{tabular}
        \end{table}

At the end of the experiment, please leave your email address and we will send you the charitable donation certificate for this experiment.

\vspace{3mm}
 \begin{center}\Large\bfseries{Instructions (Experimental Task 3)}\end{center}
 \vspace{3mm}

\noindent{\textbf{ Questionnaire 1}}

1. What is your gender?

\quad\scriptsize{\textcircled{}}\normalsize{} Male

\quad\scriptsize{\textcircled{}}\normalsize{} Female

2. What is your age range?

\quad\scriptsize{\textcircled{}}\normalsize{} 18-20 years old

\quad\scriptsize{\textcircled{}}\normalsize{} 21-23 years old

\quad\scriptsize{\textcircled{}}\normalsize{} 24-26 years old

\quad\scriptsize{\textcircled{}}\normalsize{} 27-29 years old

\quad\scriptsize{\textcircled{}}\normalsize{} Over 30 years old

3. What is the duration of your education?

\quad\scriptsize{\textcircled{}}\normalsize{} 2 years or less

\quad\scriptsize{\textcircled{}}\normalsize{} 3 years

\quad\scriptsize{\textcircled{}}\normalsize{} 4 years

\quad\scriptsize{\textcircled{}}\normalsize{} 5 years

\quad\scriptsize{\textcircled{}}\normalsize{} More than 5 years

4. What is your level of education?

\quad\scriptsize{\textcircled{}}\normalsize{} Secondary Specialized

\quad\scriptsize{\textcircled{}}\normalsize{} Associate Degree

\quad\scriptsize{\textcircled{}}\normalsize{} Bachelor's Degree

\quad\scriptsize{\textcircled{}}\normalsize{} Master's Degree

\quad\scriptsize{\textcircled{}}\normalsize{} Doctorate

5. Are you willing to invest 20\% of your annual income in a moderately growing diversified fund?

\quad\scriptsize{\textcircled{}}\normalsize{} Extremely unlikely

\quad\scriptsize{\textcircled{}}\normalsize{} Unlikely

\quad\scriptsize{\textcircled{}}\normalsize{} Neutral

\quad\scriptsize{\textcircled{}}\normalsize{} Likely

\quad\scriptsize{\textcircled{}}\normalsize{} Extremely likely

6. Are you willing to invest 10\% of your annual income in highly speculative stocks?

\quad\scriptsize{\textcircled{}}\normalsize{} Extremely unlikely

\quad\scriptsize{\textcircled{}}\normalsize{} Unlikely

\quad\scriptsize{\textcircled{}}\normalsize{} Neutral

\quad\scriptsize{\textcircled{}}\normalsize{} Likely

\quad\scriptsize{\textcircled{}}\normalsize{} Extremely likely

7. Are you willing to participate in skydiving activities?

\quad\scriptsize{\textcircled{}}\normalsize{} Extremely unlikely

\quad\scriptsize{\textcircled{}}\normalsize{} Unlikely

\quad\scriptsize{\textcircled{}}\normalsize{} Neutral

\quad\scriptsize{\textcircled{}}\normalsize{} Likely

\quad\scriptsize{\textcircled{}}\normalsize{} Extremely likely

8. Overall, I am a person who is willing to take risks.

\quad\scriptsize{\textcircled{}}\normalsize{} Extremely unlikely

\quad\scriptsize{\textcircled{}}\normalsize{} Unlikely

\quad\scriptsize{\textcircled{}}\normalsize{} Neutral

\quad\scriptsize{\textcircled{}}\normalsize{} Likely

\quad\scriptsize{\textcircled{}}\normalsize{} Extremely likely

9. If you unexpectedly received 1000 yuan today, how much would you donate to charitable causes?

10. How willing are you to donate to public welfare causes without expecting anything in return?

\quad\scriptsize{\textcircled{}}\normalsize{} Extremely unlikely

\quad\scriptsize{\textcircled{}}\normalsize{} Unlikely

\quad\scriptsize{\textcircled{}}\normalsize{} Neutral

\quad\scriptsize{\textcircled{}}\normalsize{} Likely

\quad\scriptsize{\textcircled{}}\normalsize{} Extremely likely

11. I believe people are generally well-intentioned.

\quad\scriptsize{\textcircled{}}\normalsize{} Extremely unlikely

\quad\scriptsize{\textcircled{}}\normalsize{} Unlikely

\quad\scriptsize{\textcircled{}}\normalsize{} Neutral

\quad\scriptsize{\textcircled{}}\normalsize{} Likely

\quad\scriptsize{\textcircled{}}\normalsize{} Extremely likely

\vspace{5mm}
\noindent{\textbf{Questionnaire 2: Raven test}}
\vspace{3mm}

This test task contains 6 figures as shown below example, each figure consists of 3x3 different patterns,where the last pattern is blank. Each row/column of these patterns is arranged according to certain rules.Your task is to find the best pattern to fill in the blanks.You have a total of 5 minutes to complete this part of the test. For every correct answer, you will get an extra 1 yuan.

\begin{figure}[h!]
    \centering
    \includegraphics[width=0.4\textwidth]{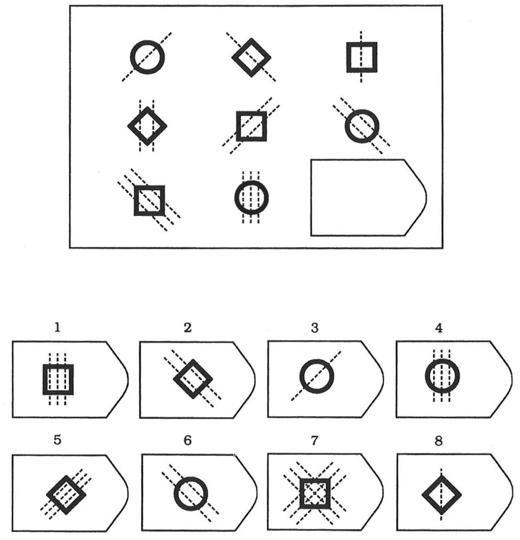}
   
    \caption{An example of a Raven task test}
\end{figure}

\vspace{5mm}
\noindent{\textbf{Questionnaire 3: CRT test}}
\vspace{3mm}

Please answer the following four questions. You have a total of 3 minutes to complete this part of the
test. You will receive an additional 1 yuan bonus for each correct answer of these questions.

1. A pair of tennis rackets and a ball together cost 1.10,and the racket is 1 more expensive than the ball. What is the price of the ball in dollars?

2. If 5 machines can produce 5 parts in 5 minutes, how many minutes would it take for 100 machines to produce 100 parts?

3. A barrel of pure water would be finished by Xiao Ming in 6 days and by Xiao Hong in 12 days. If Xiao Ming and Xiao Hong become roommates and drink from the same barrel, how many days would it take for them to finish the water?

4. As shown in the figure below (which is not provided here), there are four cards (A, B, C, D) on the table. Each card has a number on the front and a color on the back. Now, Xiao Ming has made the following conjecture: If the front of a card is an even number, then its back is blue. Assuming you can look at these cards, which cards must you turn over to verify whether Xiao Ming's conjecture is correct?
\begin{figure}[h!]
    \centering
    \includegraphics[width=0.6\textwidth]{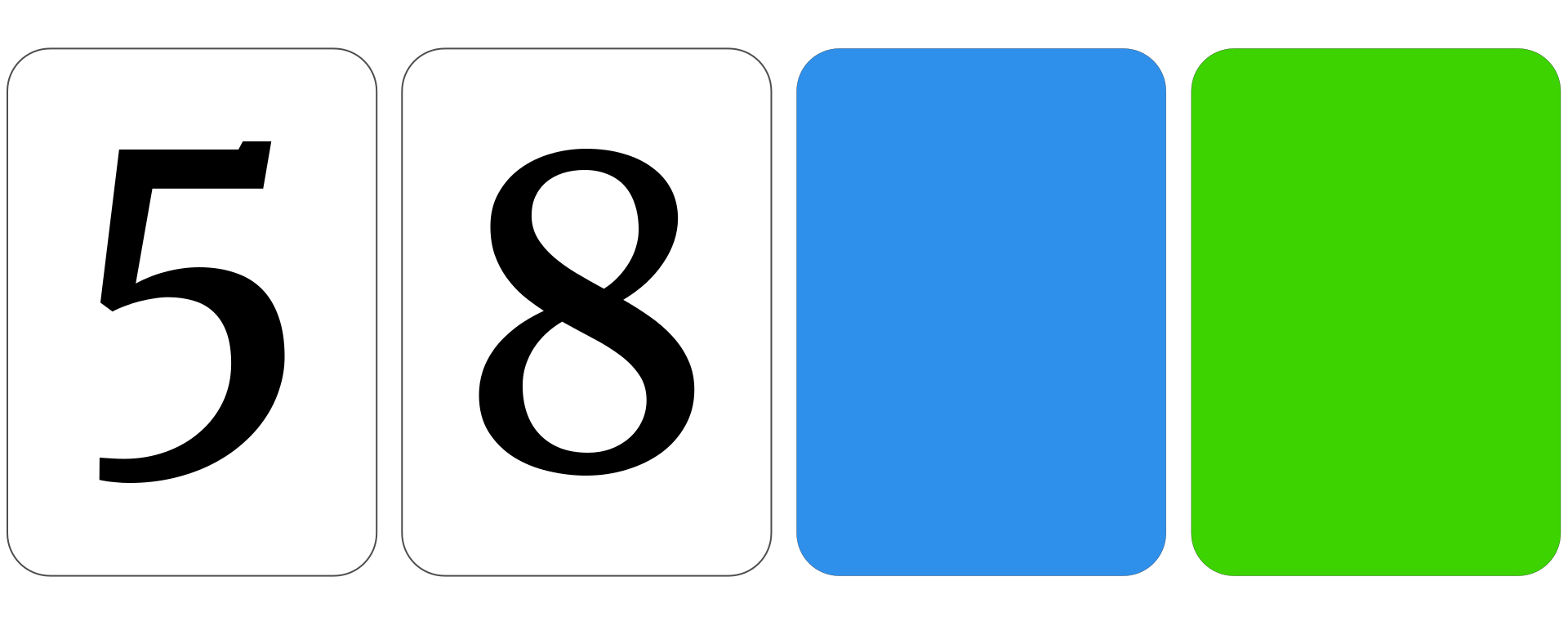}
\end{figure}

\vspace{5mm}
\noindent{\textbf{Questionnaire 4: Algorithm literacy and awareness}}
\vspace{3mm}

Below are some descriptions related to the field of artificial intelligence. For each statement, please select according to your thoughts.

1. Are you aware of the current state of development in artificial intelligence (AI)?

\scriptsize{\textcircled{}}\normalsize{}  Completely Unaware \scriptsize{\textcircled{}}\normalsize{} Not Very Aware \scriptsize{\textcircled{}}\normalsize{} Somewhat Aware \scriptsize{\textcircled{}}\normalsize{} Aware \scriptsize{\textcircled{}}\normalsize{} Very Aware

2. Are you aware of the potential use of AI algorithms in medical decision-making?

\scriptsize{\textcircled{}}\normalsize{}  Completely Unaware \scriptsize{\textcircled{}}\normalsize{} Not Very Aware \scriptsize{\textcircled{}}\normalsize{} Somewhat Aware \scriptsize{\textcircled{}}\normalsize{} Aware \scriptsize{\textcircled{}}\normalsize{} Very Aware

3. Do you think AI algorithms will have what kind of impact on the future development of the medical industry?

\scriptsize{\textcircled{}}\normalsize{} Very Negative Impact
\scriptsize{\textcircled{}}\normalsize{} Negative Impact
\scriptsize{\textcircled{}}\normalsize{} No Impact
\scriptsize{\textcircled{}}\normalsize{} Positive Impact
\scriptsize{\textcircled{}}\normalsize{} Very Positive Impact

4. Are you aware of the basic principles of machine learning?

\scriptsize{\textcircled{}}\normalsize{}  Completely Unaware \scriptsize{\textcircled{}}\normalsize{} Not Very Aware \scriptsize{\textcircled{}}\normalsize{} Somewhat Aware \scriptsize{\textcircled{}}\normalsize{} Aware \scriptsize{\textcircled{}}\normalsize{} Very Aware

5. Do you believe that a critical thinking approach should be maintained when using AI algorithms?

\scriptsize{\textcircled{}}\normalsize{} Strongly Disagree
\scriptsize{\textcircled{}}\normalsize{} Disagree
\scriptsize{\textcircled{}}\normalsize{} Neutral
\scriptsize{\textcircled{}}\normalsize{} Agree
\scriptsize{\textcircled{}}\normalsize{} Strongly Agree

\vspace{5mm}
Below are some relevant descriptions of AI for the healthcare field, for each of the following statements, please choose as you see fit.

1. How useful do you think AI algorithms are in medical decision making?

\scriptsize{\textcircled{}}\normalsize{} Almost no effect
\scriptsize{\textcircled{}}\normalsize{} Limited role
\scriptsize{\textcircled{}}\normalsize{} Somewhat useful
\scriptsize{\textcircled{}}\normalsize{} Very useful

2. How useful do you think AI algorithms are in assisting physicians' treatment systems?

\scriptsize{\textcircled{}}\normalsize{} Almost no effect
\scriptsize{\textcircled{}}\normalsize{} Limited role
\scriptsize{\textcircled{}}\normalsize{} Somewhat useful
\scriptsize{\textcircled{}}\normalsize{} Very useful

3. How useful do you think AI algorithms are in healthcare data analysis?

\scriptsize{\textcircled{}}\normalsize{} Almost no effect
\scriptsize{\textcircled{}}\normalsize{} Limited role
\scriptsize{\textcircled{}}\normalsize{} Somewhat useful
\scriptsize{\textcircled{}}\normalsize{} Very useful

4. Have you ever questioned or validated the results of AI algorithms in healthcare decision-making?

\scriptsize{\textcircled{}}\normalsize{} Not at all
\scriptsize{\textcircled{}}\normalsize{} Only once
\scriptsize{\textcircled{}}\normalsize{} A few times
\scriptsize{\textcircled{}}\normalsize{} Regularly

5. Would you be willing to receive specialized training on the use of AI in healthcare?

\scriptsize{\textcircled{}}\normalsize{} Totally unwilling
\scriptsize{\textcircled{}}\normalsize{} Not very willing
\scriptsize{\textcircled{}}\normalsize{} Willing 
\scriptsize{\textcircled{}}\normalsize{} Very willing

\vspace{5mm}
\noindent{\textbf{Questionnaire 5: Algorithm fairness and trust}}
\vspace{3mm}

1. Which AI big language model have you used?

\quad $\Square$  ChatGPT

\quad $\Square$ Gemini

\quad $\Square$ Wen Xin Yi Yan

\quad $\Square$ Kimi chat

\quad $\Square$ Sora

\quad $\Square$ Dall.E3

\quad $\Square$ Other

\quad $\Square$ Don't know

If you selected ``Other", please enter your answer below.

2. How often do you use the Big Language Model?

\scriptsize{\textcircled{}}\normalsize{} Every day
\scriptsize{\textcircled{}}\normalsize{} 3 times a week
\scriptsize{\textcircled{}}\normalsize{} Once every six months
\scriptsize{\textcircled{}}\normalsize{}Never

3. Below are two different perspectives on Al for healthcare, for each of the following statements, please choose as you see fit. Whichever side of the argument you agree with, select the scale point that is close to that argument.(There are five points on the scale, and the meanings from left to right are ``Strongly Agree With Viewpoint A", ``Somewhat Agree With Viewpoint A", ``Neutral", ``Strongly Agree with Viewpoint B", ``Strongly Agree with Viewpoint B").

(1)View A : AI technology has made it more difficult for patients in low-income or remote areas to access quality healthcare.

\ \quad View B : AI technology makes it more likely that patients in low-income or remote areas will have access to high-quality healthcare.

\qquad Strongly Agree With  A  \scriptsize{\textcircled{}}\normalsize{}----\scriptsize{\textcircled{}}\normalsize{}----\scriptsize{\textcircled{}}\normalsize{}----\scriptsize{\textcircled{}}\normalsize{}----\scriptsize{\textcircled{}}\normalsize{}  Strongly Agree With  B

(2)View A : AI technology has a significant impact on healthcare access equity.

\ \quad View B : AI technology did not have any significant impact on equity of access to healthcare.

\qquad Strongly Agree With  A  \scriptsize{\textcircled{}}\normalsize{}----\scriptsize{\textcircled{}}\normalsize{}----\scriptsize{\textcircled{}}\normalsize{}----\scriptsize{\textcircled{}}\normalsize{}----\scriptsize{\textcircled{}}\normalsize{}  Strongly Agree With  B

(3)View A : AI technology makes resources tend to be0allocated to patients or providers who can pay higher fees.

\ \quad View B : AI technology has led to a more even distribution of resources.

\qquad Strongly Agree With  A  \scriptsize{\textcircled{}}\normalsize{}----\scriptsize{\textcircled{}}\normalsize{}----\scriptsize{\textcircled{}}\normalsize{}----\scriptsize{\textcircled{}}\normalsize{}----\scriptsize{\textcircled{}}\normalsize{}  Strongly Agree With  B

(4)View A : The use of AI technology may exacerbate inequalities in research and development of effective treatments for certain diseases.

\ \quad View B : AI technology has made it possible for rare diseases to be more fully researched as well.

\qquad Strongly Agree With  A  \scriptsize{\textcircled{}}\normalsize{}----\scriptsize{\textcircled{}}\normalsize{}----\scriptsize{\textcircled{}}\normalsize{}----\scriptsize{\textcircled{}}\normalsize{}----\scriptsize{\textcircled{}}\normalsize{}  Strongly Agree With  B

(5)View A : AI technology may cause physicians to become overly reliant on technology at the expense of direct patient interaction.

\ \quad View B : AI technology may increase patient trust in healthcare because it provides more accurate medical information.

\qquad Strongly Agree With  A  \scriptsize{\textcircled{}}\normalsize{}----\scriptsize{\textcircled{}}\normalsize{}----\scriptsize{\textcircled{}}\normalsize{}----\scriptsize{\textcircled{}}\normalsize{}----\scriptsize{\textcircled{}}\normalsize{}  Strongly Agree With  B

(6)View A : AI provides inaccurate results.

\ \quad View B : AI provides accurate results.

\qquad Strongly Agree With  A  \scriptsize{\textcircled{}}\normalsize{}----\scriptsize{\textcircled{}}\normalsize{}----\scriptsize{\textcircled{}}\normalsize{}----\scriptsize{\textcircled{}}\normalsize{}----\scriptsize{\textcircled{}}\normalsize{}  Strongly Agree With  B

(7)View A : AI provides results that are not easily applied to common problems.

\ \quad View B : AI provides enough results to apply to common problems.

\qquad Strongly Agree With  A  \scriptsize{\textcircled{}}\normalsize{}----\scriptsize{\textcircled{}}\normalsize{}----\scriptsize{\textcircled{}}\normalsize{}----\scriptsize{\textcircled{}}\normalsize{}----\scriptsize{\textcircled{}}\normalsize{}  Strongly Agree With  B

(8)View A : If hospitals introduce AI-assisted medical technology, I'm not so sure about using AI for assisted diagnosis and treatment.

\ \quad View B : If the hospital introduces AI-assisted healthcare technology, I will combine it with a big model to assist in the consultation.

\qquad Strongly Agree With  A  \scriptsize{\textcircled{}}\normalsize{}----\scriptsize{\textcircled{}}\normalsize{}----\scriptsize{\textcircled{}}\normalsize{}----\scriptsize{\textcircled{}}\normalsize{}----\scriptsize{\textcircled{}}\normalsize{}  Strongly Agree With  B

(9)View A : Using the Large Language Model will not improve the accuracy of a doctor's diagnosis.

\ \quad View B : Using the Large Language Model can improve the accuracy of a doctor's diagnosis.

\qquad Strongly Agree With  A  \scriptsize{\textcircled{}}\normalsize{}----\scriptsize{\textcircled{}}\normalsize{}----\scriptsize{\textcircled{}}\normalsize{}----\scriptsize{\textcircled{}}\normalsize{}----\scriptsize{\textcircled{}}\normalsize{}  Strongly Agree With  B

(10)View A : AI algorithms have a very negative impact on the future of the healthcare industry.

\quad\ \, View B : AI algorithms have a very positive impact on the future of the healthcare industry.

\qquad Strongly Agree With  A  \scriptsize{\textcircled{}}\normalsize{}----\scriptsize{\textcircled{}}\normalsize{}----\scriptsize{\textcircled{}}\normalsize{}----\scriptsize{\textcircled{}}\normalsize{}----\scriptsize{\textcircled{}}\normalsize{}  Strongly Agree With  B

\subsection{Screenshots}

\begin{figure}[h!]
    \centering
    \includegraphics[width=1\textwidth]{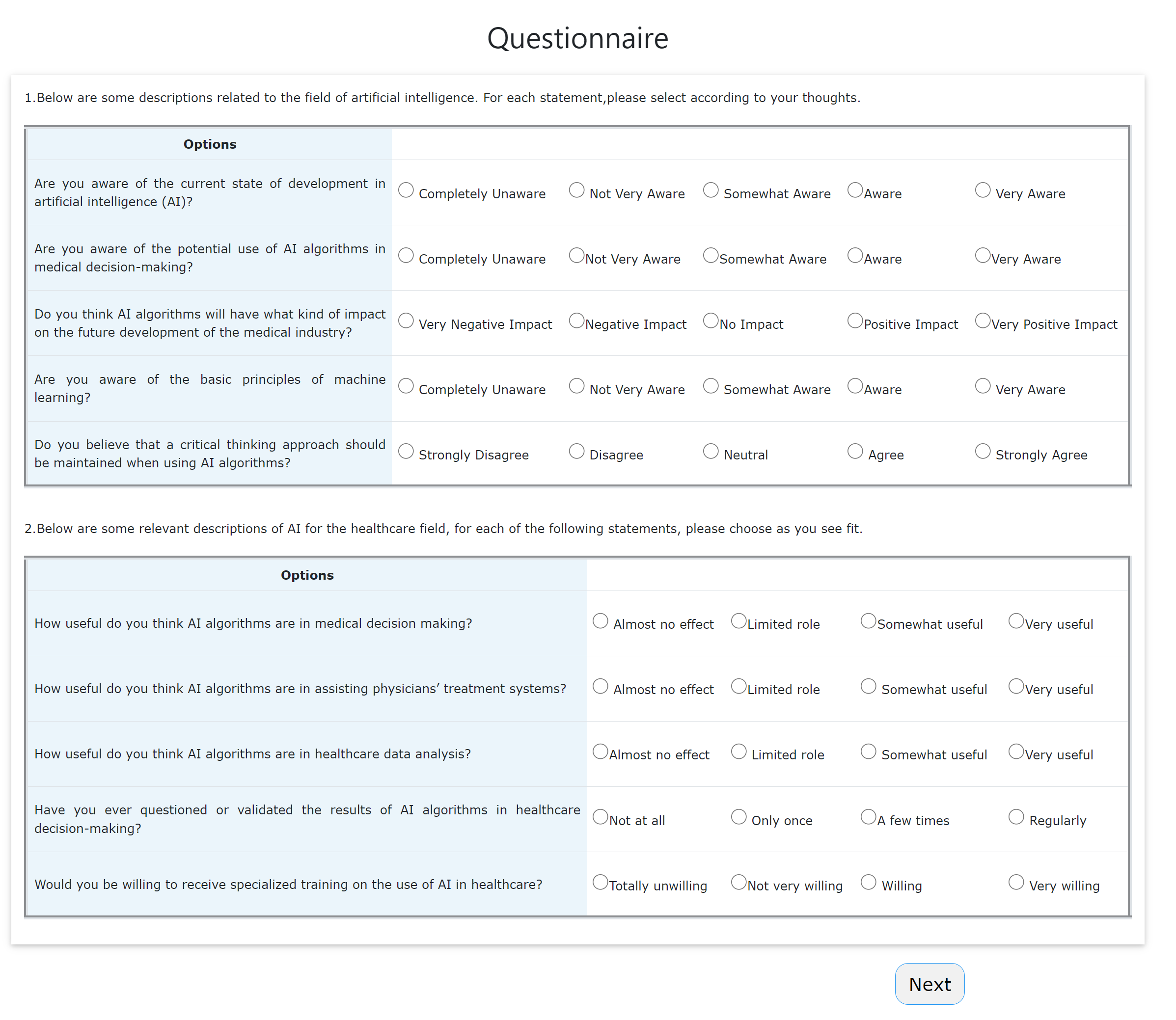}
    
\end{figure}
\begin{figure}[h!]
    \centering
    \includegraphics[width=1\textwidth]{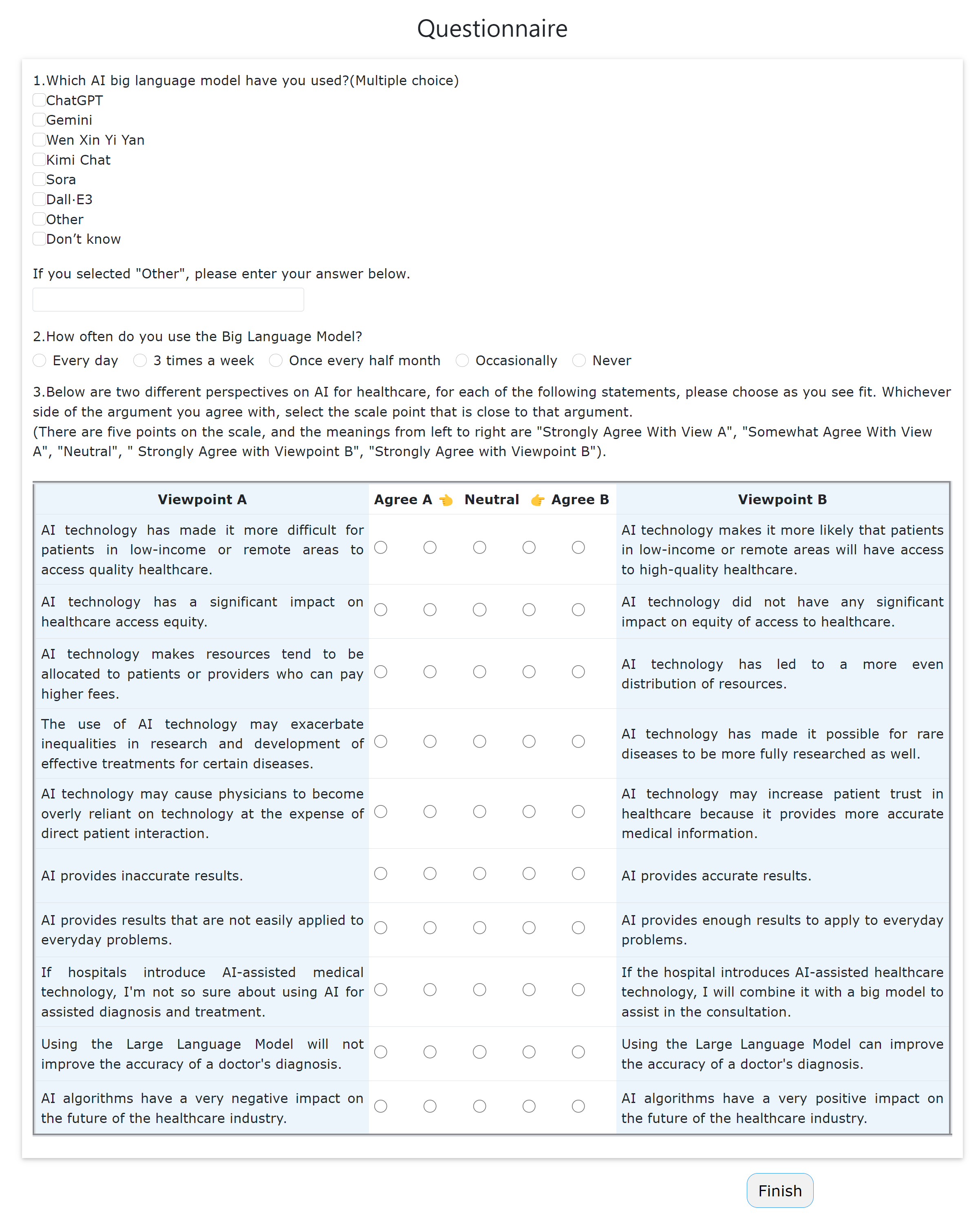}
    
\end{figure}

\newpage

\end{appendices}

\end{document}